%% file: nim.tex
\documentclass{elsart}

\usepackage{epsfig}
\usepackage{amsmath,bm}
\usepackage{amscd}
\usepackage{amssymb}
\usepackage{floatflt}
\usepackage{rotating}
\usepackage{wrapfig} 
\input{abbrevia}

\begin{document}
\begin{frontmatter}
\bibliographystyle{elsart-num} 
 \title{Energy Linearity and Resolution of the 
        ATLAS Electromagnetic Barrel Calorimeter in an Electron Test-Beam
  }

%\author[Bla]{ATLAS Electromagnetic Barrel Calorimeter Collaboration}
%
\author[Annecy]{M.~Aharrouche},
\author[Annecy]{J.~Colas},
\author[Annecy]{L.~Di Ciaccio},
\author[Annecy]{M.~El~Kacimi\thanksref{ElKacimi}},
\author[Annecy]{O.~Gaumer},
\author[Annecy]{M.~Gouan\`ere},
\author[Annecy]{D.~Goujdami\thanksref{ElKacimi}},
\author[Annecy]{R.~Lafaye},
\author[Annecy]{S.~Laplace},
\author[Annecy]{C.~Le Maner},
\author[Annecy]{L.~Neukermans},
\author[Annecy]{P.~Perrodo},
\author[Annecy]{L.~Poggioli},
\author[Annecy]{D.~Prieur},
\author[Annecy]{H.~Przysiezniak},
\author[Annecy]{G.~Sauvage},
\author[Annecy]{F.~Tarrade},
\author[Annecy]{I.~Wingerter-Seez},
\author[Annecy]{R.~Zitoun},
\author[Brookhaven]{F.~Lanni},
\author[Brookhaven]{H.~Ma},
\author[Brookhaven]{S.~Rajagopalan},
\author[Brookhaven]{S.~Rescia},
\author[Brookhaven]{H.~Takai},
\author[Casablanca]{A.~Belymam},
\author[Casablanca]{D.~Benchekroun},
\author[Casablanca]{M.~Hakimi},
\author[Casablanca]{A.~Hoummada},
\author[Dallas]{E.~Barberio\thanksref{Barberio}},
\author[Dallas]{Y.S.~Gao},
\author[Dallas]{L.~Lu},
\author[Dallas]{R. Stroynowski},
\author[CERN]{M.~Aleksa},
\author[CERN]{J.~Beck Hansen\thanksref{Beck}},
\author[CERN]{T.~Carli\thanksref{cauthor}},
\author[CERN]{I.~Efthymiopoulos},
\author[CERN]{P.~Fassnacht},
\author[CERN]{F.~Follin},
\author[CERN]{F.~Gianotti},
\author[CERN]{L.~Hervas},
\author[CERN]{W.~Lampl},
\author[Grenoble]{J.~Collot},
\author[Grenoble]{J.Y.~Hostachy},
\author[Grenoble]{F.~Ledroit-Guillon},
\author[Grenoble]{P.~Martin},
\author[Grenoble]{F.~Ohlsson-Malek},
\author[Grenoble]{S. Saboumazrag},
\author[Nevis]{M.~Leltchouk},
\author[Nevis]{J.A.~Parsons},
\author[Nevis]{M.~Seman},
\author[Nevis]{S.~Simion},
\author[Milano]{D.~Banfi}
\author[Milano]{L.~Carminati},
\author[Milano]{D.~Cavalli},
\author[Milano]{G.~ Costa},
\author[Milano]{M.~Delmastro},
\author[Milano]{M.~Fanti},
\author[Milano]{L.~Mandelli},
\author[Milano]{M.~Mazzanti},
\author[Milano]{G.~F.~Tartarelli},
\author[Orsay]{C.~Bourdarios},
\author[Orsay]{L.~Fayard},
\author[Orsay]{D.~Fournier},
\author[Orsay]{G.~Graziani},
\author[Orsay]{S.~Hassani},
\author[Orsay]{L.~Iconomidou-Fayard},
\author[Orsay]{M.~Kado},
\author[Orsay]{M.~Lechowski},
\author[Orsay]{M.~Lelas},
\author[Orsay]{G.~Parrour},
\author[Orsay]{P.~Puzo},
\author[Orsay]{D.~Rousseau},
\author[Orsay]{R.~Sacco\thanksref{Sacco}},
\author[Orsay]{L.~Serin},
\author[Orsay]{G.~Unal},
\author[Orsay]{D.~Zerwas},
\author[Jussieu]{A.~Camard},
\author[Jussieu]{D.~Lacour},
\author[Jussieu]{B.~Laforge},
\author[Jussieu]{I.~Nikolic-Audit},
\author[Jussieu]{Ph.~Schwemling},
\author[Rabat1]{H.~Ghazlane},
\author[Rabat]{R.~Cherkaoui El Moursli},
\author[Rabat]{A.~Idrissi Fakhr-Eddine},
\author[Saclay]{M.~Boonekamp},
\author[Saclay]{N.~Kerschen},
\author[Saclay]{B.~Mansouli\'{e}},
\author[Saclay]{P.~Meyer},
\author[Saclay]{J.~Schwindling},
\author[Stockholm]{B.~Lund-Jensen},
\author[Stockholm]{Y.~Tayalati}
\address[Annecy]{Laboratoire de Physique de Particules (LAPP),
IN2P3-CNRS, F-74941~Annecy-le-Vieux~Cedex, France.}
\address[Brookhaven]{Brookhaven National Laboratory (BNL), Upton,
  NY~11973-5000, USA.}
\address[Casablanca]{Facult\'{e} des Sciences A\"{\i}n Chock, Casablanca,
  Morocco.}
\address[Dallas]{Southern Methodist University, Dallas, Texas 75275-0175,
  USA.}
\address[CERN]{European Laboratory for Particle Physics (CERN),
  CH-1211~Geneva~23, Switzerland.}
\address[Grenoble]{Laboratoire de Physique Subatomique et de Cosmologie,
  Universit\'e Joseph Fourier, IN2P3-CNRS, F-38026~Grenoble, France.}
\address[Nevis]{Nevis Laboratories, Columbia University, Irvington,
  NY~10533, USA.}
\address[Milano]{Dipartimento di Fisica dell'Universit\`{a} di Milano and
  INFN, I-20133~Milano, Italy.}
\address[Orsay]{Laboratoire de l'Acc\'{e}l\'{e}rateur Lin\'{e}aire,
  Universit\'{e} de Paris-Sud, IN2P3-CNRS, F-91898~Orsay~Cedex, France.}
\address[Jussieu]{Universit\'es Paris VI et VII, Laboratoire de Physique
  Nucl\'eaire et de Hautes Energies, F-75252 Paris, France.}
\address[Rabat1]{Facult\'e des Sciences and
  Centre National de l'\'Energie des Sciences et des Techniques
  Nucl\'eaires, Rabat, Morocco.}
\address[Rabat]{Universit\'e Mohamed V, Facult\'e des Sciences, Rabat, Morocco.}
\address[Saclay]{CEA, DAPNIA/Service de Physique des Particules,
  CE-Saclay, F-91191~Gif-sur-Yvette~Cedex, France.}
\address[Stockholm]{Royal Institute of Technology, Stockholm, Sweden.}

\thanks[Barberio]{Now at university of Melbourne, Australia.}
\thanks[ElKacimi]{Visitor from LPHEA, FSSM-Marrakech (Morroco).}
\thanks[Beck]{Now at Niels Bohr Institute, Copenhagen.}
%\thanks[Deceased]{Deceased.}
\thanks[cauthor]{E-mail: Tancredi.Carli@cern.ch.}
\thanks[Sacco]{Now at Queen Mary, University of London.}

\begin{abstract}
A module of the ATLAS electromagnetic barrel
liquid argon calorimeter was exposed %at $\eta=0.687$ 
to the CERN electron test-beam
at the H8 beam line upgraded for precision momentum measurement.
The available energies of the electron beam ranged from $10$ to $245$\GeVx.  
The electron beam impinged at one point corresponding to a pseudo-rapidity
of $\eta=0.687$ and an azimuthal angle of $\phi=0.28$ in the ATLAS coordinate system.
A detailed study of several effects biasing the
electron energy measurement allowed an energy reconstruction procedure 
to be developed that ensures a good linearity and a good resolution.
Use is made of detailed Monte Carlo simulations based on  
\Geant which describe the longitudinal and transverse
shower profiles as well as the energy distributions.
For electron energies between $15$\GeV and $180$\GeV the deviation of the measured incident electron
energy over the  beam energy is within $0.1\%$. 
The systematic uncertainty of the measurement is about $0.1\%$ % per effect
at low energies and negligible at high energies.
The energy
resolution is found to be about $10$\% $\cdot \sqrt{E}$ for the sampling term 
and about $0.2$\% for the local constant term.
\end{abstract}
%
%
%\bigskip
%\bigskip
\end{frontmatter}

%\end{titlepage}

%\tableofcontents
%\newpage

\section*{Introduction}
\noindent 
The Large Hadron Collider (LHC), currently under construction at CERN,
will collide protons on protons with a beam energy of $7$~{\rm TeV},
extending the available centre-of-mass energy %($\sqrt{s}$) 
by about an order of magnitude over that of existing colliders. Together with its high
collision rate, corresponding to an expected integrated luminosity of
$10-100~{\mathrm{ fb}}^{-1}/{\mathrm{ year}}$, these energies allow for production
of particles with high masses or high transverse momenta or other processes with low production
cross-sections. The LHC will search for effects of new interactions at very short
distances and for new particles beyond the Standard Model of
particle physics (SM).
The large particle production rates at LHC are not only a challenge to our
theoretical understanding of proton proton collisions at such high energies,
but also for the detectors.

An excellent knowledge of the electron or photon energy 
is needed for precision measurements of, for example couplings within and beyond the SM,
or to resolve possible narrow resonances of new particles over a large background. 
A good energy resolution and a good linearity need to be achieved for energies
ranging from a few\GeV up to a few\TeVx. 

A prominent example is the possible discovery of the Higgs boson which in the SM
provides an explanation how the elementary particles acquire mass.
If the Higgs boson mass is below $130$\GeVx,
the decay $H\rightarrow \gamma\gamma$ is
the most promising discovery channel. If the Higgs mass is larger and, in particular if
it is at least twice the mass of the $Z^0$-boson $2\,M_Z \sim 180$\GeVx, 
the Higgs boson can be discovered in the $H \rightarrow Z^0 Z^0 \rightarrow e^+ e^- e^+ e^-$
decay channel. Even in this case 
the energy of one of the electrons can be as low as about $10$\GeVx.
The possible observation of the Higgs boson
requires therefore  excellent measurements of electrons and photons from low
to high energies.

The absolute energy measurement can be calibrated on reference reactions as
$p p \to Z^0 X  \to e^+ e^- X$, exploiting the precise knowledge of the mass of the
$Z^0$-boson. However, a good energy resolution and a good linearity can only be achieved,
if the detector, the physics processes in the detector, and effects of the
read-out electronics are well understood. In particular, knowledge of the
detector linearity determines how precisely an energy measurement at one
particular energy can be transfered to any energy. For instance, to measure
the mass of the  $W^\pm$-boson with a precision of $15$\MeV  
a linearity of about $10^{-4}$ is required in an energy interval
which is given by the difference of the transverse energy spectrum of an electron
from the $W^\pm$-boson decay and that of the  $Z^0$-boson~\cite{TDRPhys}.

The electromagnetic (EM) barrel liquid argon (\LArx) calorimeter 
is the main detector to measure the electron energy 
in the central part of the ATLAS detector.
It is a sampling calorimeter with accordion shaped
lead absorbers and \LAr as active medium. 
%It covers the central rapidity region, i.e. $|\eta|<1.42$.
%In addition, it is equally important that the data are well described
%by Monte Carlo simulations.

In August $2002$ a production module of the ATLAS \LAr EM
barrel calorimeter  was exposed to an electron beam in the
energy range of $10$ to $245$\GeV at the CERN H8 beam line, which was
upgraded with a system to precisely measure the beam energy.  
These data are used to assess the linearity of the electron 
energy measurement and the energy resolution.
A calibration scheme is developed
which ensures simultaneously a good linearity and a good resolution.

In the past the linearity of the ATLAS EM calorimeter 
has been studied with a
calorimeter prototype \cite{TDR}. For electron energies between
$20$ - $300$\GeV a linearity within $1$\% has been measured.

In section~\ref{sec:beam} the system to measure the linearity of the beam energy
is presented and its accuracy is discussed.
Section~\ref{sec:set-up} describes the ATLAS EM barrel calorimeter,
the H8 test-beam set-up, the data samples, and the event selection.
The Monte Carlo simulation is out-lined in section~\ref{sec:g4}.
Section~\ref{sec:calib} summarises the calibration of the electronic signal, converting
the measured current to a visible energy, i.e., the energy deposited in the active medium.
Section~\ref{sec:calibration_emshower} discusses general effects related to
the physics of EM showers that need to be taken into account to precisely
reconstruct the electron energy.
Section~\ref{sec:reco} presents the calibration procedure to precisely reconstruct
the total electron energy.
Comparison of the visible energies and the total reconstructed energy
in the data and in the Monte Carlo simulation are shown in section~\ref{sec:data_mc_comparision}.
The possible pion contamination in the electron beam is discussed 
in section~\ref{sec:pion_contamination}.
The results of the energy measurement together with their systematic uncertainties
are presented in section~\ref{sec:results}.

%\newpage
\section{Precise Determination of the Relative Electron Beam Energy}
\label{sec:beam}
\subsection{The H8 Beam-line}
\label{sec:beamsetup}
\begin{figure}[th]
\begin{center}
\psfig{figure=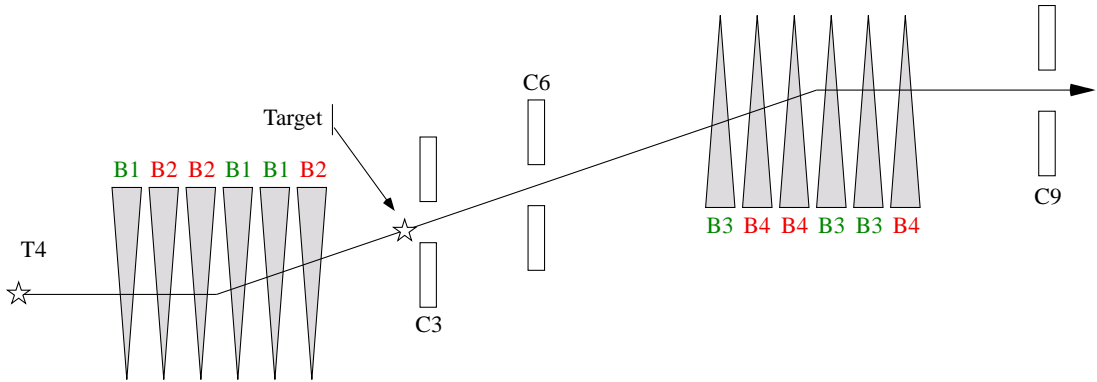,
%bbllx=16,bblly=50,bburx=530,bbury=630,angle=270,clip,
width=14.5cm}
\end{center}
%\vspace{-1.cm}
\caption
{Sketch of the H8 beam line showing the momentum definition elements.
 The magnets labeled B4 were degaussed during the data taking period.
\label{fig:beamline}}
\end{figure}

\noindent 
The H8 beam line %\cite{h8beamline} 
is sketched in Fig.~\ref{fig:beamline}. 
The  electron momentum definition %used in the work reported below, 
is based on the second
momentum analysis using two triplets of bending magnets B3 and B4, between collimator C3
acting as a source, and collimator C9 acting as a momentum slit, while the upstream part 
was set at  $180$\GeVx.
A thin sheet of lead was introduced upstream of C3 to increase
the electron yield.
%At all energies the downstream part of the beam line was setup first, then
%the upstream part was tuned so as to reach a maximum rate at C9.
The magnets were set in direct current (DC) mode and the induced current was read-out 
with a Direct Current Current Transformer (DCCT).

To control the  induced current in the spectrometer with a single
precision DCCT, only the B3 magnets,
connected in series, were used. This limited the maximum momentum
to $180$\GeV for a current of about $1200$~{\rm A}. 
The B4 magnets were degaussed, following a bipolar loop and kept
unpowered. The remaining bending power $\int B dl$ was zero 
with an uncertainty of $\pm 1.5$~{\rm mT~$\cdot$~m}. 
Each bending magnet has an effective length of $5.2$~{\rm m} 
and a field of $1.42$~{\rm T/kA} (linear part).

To eliminate any uncertainty coming from the geometry
of the spectrometer, the jaw positions of both C3 and C9 were kept
fixed during the entire data taking period. A slit of $8$~{\rm mm} was chosen for C3
as a compromise between the beam intensity and the momentum spread. %, i.e. accuracy and rate. 
The slit of C9 was kept at $8$~{\rm mm}. 
The induced momentum spread was $\pm 0.15$~\% at all energies.
With the geometry of the spectrometer fixed (its total deviation
is $41$~{\rm mrad}) the momentum of selected electrons is directly proportional
to the bending power $\int B dl$ of the B3 triplet. A correction of half the
energy lost by synchrotron radiation in the spectrometer 
was applied to particles at the detector, downstream of C9.

The effect of the earth's field along the beam line was also evaluated.
%While a naive calculation gives a positive shift of $22$\MeV on negative particles, 
%The effect of the quadrupoles in the beam-line simulated
%with the TURTLE program\cite{turtle} has been found to shift the electron energy by
%$4$\MeVx. The shift is positive for negative particles.
%        
Taking into account the focusing effect of the quadrupoles with the 
beam transport program TURTLE\cite{turtle},
the net effect found was a shift of $+4$\MeV for negative particles.

%%%%%%%%%%%%%%%%%%%%%%%%%%%%%%%%%%%%%%%%%%%%%%%%%%%%%%%%%
%%%%%%%%%%%%%%%%%%%%%%%%%%%%%%%%%%%%%%%%%%%%%%%%%%%%%%%%%%%%%%%%%g

\subsection{Control of the Beam Energy Linearity}
\label{sec:beamlinearity}
\begin{figure}[th]
\begin{center}
\psfig{figure=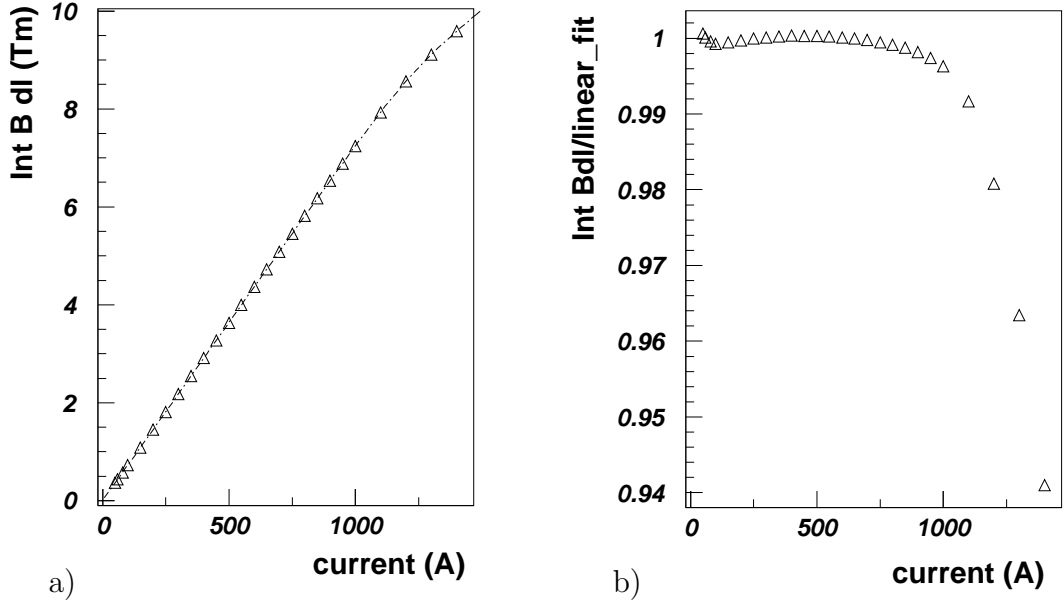,
%bbllx=16,bblly=50,bburx=530,bbury=630,angle=0,clip,
width=14.cm}
\end{center}
\begin{picture}(0,0) 
\put( 5, 5){a)}
\put(80, 5){b)}
\end{picture}
\vspace{-0.5cm}
\caption
{a) Calibration measurements of the magnetic field 
 integrated over the beam path as a function of the induced current.
 The adjusted parameterisation is superimposed as line.
 b) Deviation from linearity, i.e., measured field values divided by a linear parameterisation
    obtained from the points below $I < 500$ {\rm A}    
    as a function of the induced current.
\label{fig:mag_field}}
\end{figure}

\noindent 
Two main sources of uncertainties on the power of the
bending magnets had to be controlled:
\begin{enumerate}
\item The value of the current in the magnet string:  \\
     The precision supply and the DCCT read-out ensured a relative precision and reproducibility of 
     $1 \cdot 10^{-4}$ \cite{iliasres,gustavson,hassani,gg}.

\item The calibration and reproducibility of the hysteresis curve: \\
     At the maximum current of $1200$~{\rm A}, the integral bending power is about
     $2$\% below the linear extrapolation from low currents (see Fig.~\ref{fig:mag_field}).
     This needed to be calibrated and the non-linearity controlled to 
     about one percent. %  of itself.
\end{enumerate}     

A reference magnet (MBN25) was %once more 
calibrated using a precision power
supply and the DCCT, and a two wire loop to measure its bending power.
The magnetic field at the centre was also measured 
with a relative precision better than $ 1 \cdot 10^{-5}$
using a set of NMR probes \cite{iliasres,gustavson} 
Results are shown in Fig.~\ref{fig:mag_field}. 
The measurements can be fitted with a polynomial function. The
residuals are smaller than $10^{-4}$.

To transport this calibration to the B3 triplet,
calibration curves measured during the time of production of about $100$ MBN
magnets were used to compare the reference magnet and the magnets of the B3 triplet.
While all magnets had been trimmed during production to be identical
within  $\pm 2 \cdot 10^{-4}$ \cite{asner,loas}, a small difference 
(at most $3 \cdot 10^{-4}$ at the highest current) 
between the reference magnet and the B3 triplet had to be corrected.

To ensure reproducibility of the field for a given current,
the same unipolar setting loop was always used, both in the bench test and
during setting up with the beam. With this procedure, 
the uncertainty on the bending
power is $1$~{\rm mT~$\cdot$~m} at all energies.

In order to have  a further cross-check of the actual field 
in the B3 triplet during electron data taking, 
one of the magnets of the triplet was instrumented with two sets
of Hall probes, to be read-out during each burst.
%The Hall probes and their read-out were borrowed from a set prepared by
%NIKHEF for the instrumentation of the ATLAS muon spectrometer \cite{atlasmuon}.
They were positioned at $1.0$ {\rm m} and $1.5$ {\rm m} inside 
the magnet, within a few~{\rm mm} from the vacuum pipe.
The Hall probes data include Hall voltages for three orthogonal directions,
and the temperature. A correction of about $-3 \cdot 10^{-4}/^oC$ 
for the magnetic field as measured with the Hall probes was applied.
By running the Hall probes positioned  in the reference magnet at the
same location as in the B3 magnets, a cross calibration with respect to
the current in the DCCT, the magnetic field at centre, 
and the bending power was obtained.

\begin{figure}[th]
\vspace{0.5cm}
\begin{center}
\psfig{figure=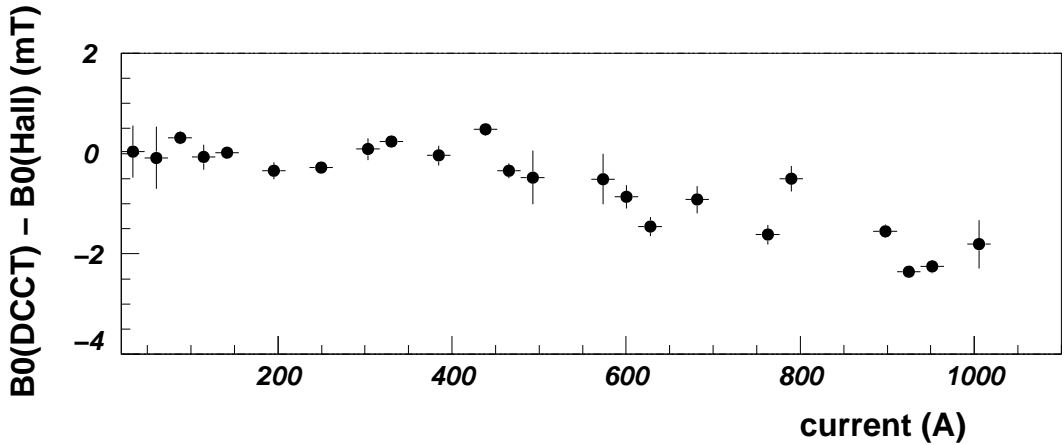,width=14.cm}
\end{center}
\vspace{-0.3cm}
\caption
{Difference between the values of the magnetic field at the magnet centre,
predicted by the calibration obtained from the current measurements and
the magnetic field measurements with the Hall probe as a function
of the current. The error bars represent the root mean square of the distribution.
\label{fig:hall}}
\end{figure}

A critical test of the cross calibration of the two field
measurements in B3 is a comparison of the magnetic field at the magnet centre calculated from
the DCCT current and from the Hall probe signals. Fig.~\ref{fig:hall} shows an excellent agreement
up to $400$ {\rm A} and a small systematic inhomogeneity ($0.15$\%) at the maximum
current. This difference is attributed to a slight difference of the
field at the Hall probe location, which is not taken into account
by the comparative calibration. A linear interpolation of the differences
leaves an average dispersion of $2 \cdot 10^{-4}$ which indicates the level
of uncertainty on the linearity induced by taking one measurement or the
other. %{\it to be checked XXXX}.

\subsection{Results and Uncertainties}
\label{sec:beamresults}
\noindent 
%For comparisons with the electron energy reconstructed in the 
%calorimeter the bending power calculated from the DCCT current was used. 
For the final comparison of the beam energy with the electron energy reconstructed in the
calorimeter, the bending power calculated from the DCCT current was used. 
For each run the DCCT currents read-out at each burst 
were averaged. The currents were stable within $0.01$~{\rm A}. 

The resultant beam energy determinations are summarised in Tab.~\ref{tab:beam_momentum}.
The synchrotron radiation correction includes small additional losses in the correction 
magnets (B5 and B6) downstream of C9.

%Since we didn't aimed in this work at a precise absolute calibration which is 
%known to about $1$\% from the beam setting, 
%the energy was arbitrarily normalised at $100$~\GeV. 
Since this work does not aim at a precise absolute calibration\footnote{The absolute beam energy
is known to about $1\%$.},
the electron beam energy was arbitrarily normalised at $100$\GeVx. 
The maximum induced uncertainty on the synchrotron radiation loss is $15$\MeV at $180$\GeVx.             
The magnet correction corresponds to the difference between MBN25
and the three magnets of the B3 string.
          
The largest uncertainty is associated to the remnant field
in B3 ($\pm 1$ {\rm mT $\cdot$ m} randomly on each energy point).
The resulting uncertainty on the linearity measurement estimated with a simple
Monte Carlo simulation where a perfect linearity is assumed and the beam momenta
fluctuate according to their uncertainties is $3 \cdot 10^{-4}$. 
A further systematic
uncertainty of $11$~\MeV due to the remnant fields in the B4 magnets must be added. 
This last uncertainty is common to all data points.

%More details on the measurement of the beam momentum can be found in ref.~\cite{gg,hassani}.

%%%%%%%%%%%%%%%%%%%%%%%%%%%%%%%%%%%%%%%%%%%%%%%%%%%%%%%%%
%%%%%%%%%%%%%%%%%%%%%%%%%%%%%%%%%%%%%%%%%%%%%%%%%%%%%%%%%%%%%%%%%g

%%%%%%%%%%%%%%%%%%%%%%%%%%%%%%%%%%%%%%%%%%%%%%%%%%%%%%%%%
%%%%%%%%%%%%%%%%%%%%%%%%%%%%%%%%%%%%%%%%%%%%%%%%%%%%%%%%%%%%%%%%%g

\input table

%\newpage
\section{Test-Beam Set-up and Event Selection}
\label{sec:set-up}
\subsection{Test-Beam Set-up}
\label{sec:test-beam-set-up}
\begin{figure}[th]
%\vspace{0.5cm}
\begin{center}
\psfig{figure=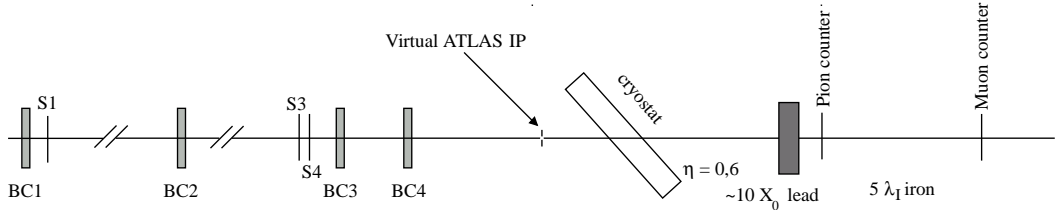,width=14.cm}
\end{center}
%\vspace{-0.3cm}
\caption
{Schematic top view of the test-beam set-up.
\label{fig:testbeamsetup}}
\end{figure}

\noindent 
The H8 test-beam set-up is shown in Fig.~\ref{fig:testbeamsetup}.
The EM barrel calorimeter is located in a cryostat filled with liquid argon (\LArx). 
The cryostat consists of an inner and an outer aluminum wall with thicknesses 
of $4.1$\cm and $3.9$\cmx, respectively.
The two walls are separated by a vacuum gap.  
Between the cryostat and the calorimeter module a foam block (ROHACELL) is installed to exclude
\LAr in front of the calorimeter.

The cryostat is mounted on a
table that allows rotation of the calorimeter in the two directions orthogonal
to the beam axis. The two directions are %conventionally 
chosen to be the
 $\eta$ and $\phi$ directions with respect to a reference frame with cylindrical 
coordinates having its origin in the virtual proton-proton interaction point
in ATLAS (see Fig.~\ref{fig:testbeamsetup}).
In this coordinate system the $z$-axis is defined along the beam axis.
The $\phi$ and $\theta$ angles are the azimuthal and polar angles. The pseudo-rapidity
is defined by $\eta = - \log{\tan{\theta/2}}$.
%
%For the runs analysed here, the table was fixed at $\eta=0.687$ and $\phi=0.28$.

In front of the cryostat
four multi-wire proportional chambers (BC1, BC2, BC3, BC4)
measured the position of the beam particles. %In between the MWPC 
Three scintillator counters (S1,S3,S4) located in between the wire chambers
were used as event trigger. The last two (S3 and S4) each with a size of $4$x$4$\cm 
were used to define the beam acceptance and to reject events with more than
one charged track.
Since in the test-beam particles hit the calorimeter at random times with respect to the
$40$~{\rm MHz} clock used by the front-end electronics, the time between
a trigger and the next clock cycle was measured with a Time Digital Converter (TDC)
with a  $50$~{\rm ps}/TDC-count sensitivity.

Behind the calorimeter and after about $13$\Xzero of material
(including cryostat and a $5$\cm lead plate) 
a scintillator was installed to reject pions (pion counter).
Another scintillator was installed after an iron block of $5$ interaction
lengths to reject muons. For most of the runs both scintillators have been used 
on-line to reject muons and pions.

%%%%%%%%%%%%%%%%%%%%%%%%%%%%%%%%%%%%%%%%%%%%%%%%%%%%%%%%%
%%%%%%%%%%%%%%%%%%%%%%%%%%%%%%%%%%%%%%%%%%%%%%%%%%%%%%%%%%%%%%%%%g

\subsection{The ATLAS Electromagnetic Barrel Calorimeter}
\label{sec:em_barrel}
\noindent 
The details of the ATLAS \LAr barrel calorimeter are described elsewhere \cite{TDR,newNIM}.
A module is made out of accordion shaped lead absorbers glued
between two $0.2\mm$thick stainless steel sheets
placed into a cryostat containing \LArx.
%In the middle between 
The read-out electrodes are interleaved between two absorbers.
At $\eta=0.687$, the lead of the absorbers have a thickness of $1.53\mm$ 
and the gap size is about $2.1\mm$on each side of the electrode.

The module is longitudinally segmented into three compartments, each having a different
transverse segmentation. At $\eta=0.687$, the front, middle and back compartments have thicknesses of 
$4.6$ \Xzerox, $17.6$\Xzero and $5.0$\Xzerox, respectively. 
The front compartment is finely segmented in
$\eta$ strips with a granularity of $0.025/8$ $\eta$-units, 
but has only four segments in $\phi$ with a granularity of $2\pi/64$.
The middle compartment has a segmentation of $0.025$ in $\eta$ and $2\pi/256$
in $\phi$. The back compartment 
has in $\phi$ the same granularity as the middle compartment, but is twice as 
coarse in  $\eta$  ($0.05$).

A thin presampler detector (PS) is mounted in front of the accordion module.
The PS consists of two straight sectors with cathode and anode electrodes glued between
plates made of a fibreglass epoxy composite (FR4).
The $13$\mm long
electrodes are oriented at a small angle with respect to the line where a particle from the test-beam
or from the nominal interaction point in ATLAS is expected to impinge on the calorimeter. 
The gap between the electrodes is $1.93$\mmx. The presampler is segmented with 
a fine granularity in $\eta$ of about $0.025$. It has four segments in  $\phi$
with a granularity of $2\pi/64$.

Between the PS and the first compartment (depending on $\phi$) 
read-out cables and electronics
like the summing- and mother-boards are installed. 
%Details can be found in \cite{perrodo}.
% can we add a reference to the construction paper (ask Isabelle suggests)

In total a full module, including the PS, has $3424$ read-out cells.

\subsection{Data Samples and Event Selection}
\label{sec:data_samples}
\noindent 
Runs at $18$ different energies between $10$\GeV and  $245$\GeV were recorded with a
ATLAS \LAr barrel calorimeter module beginning of August $2002$ 
within three days. Approximately every 12 hours calibration
runs were taken. Some of the runs were repeated with the same beam energy at different
times during the data taking period. No systematic effect was found.
The temperature variation of the \LAr was within $7$~{\rm mK} over the total $2002$ running period, which corresponds to
a maximum variation of the calorimeter response of $\pm 7 \cdot 10^{-5}$.

The electron beam impinged on the module at an angle corresponding to a virtual angle in the
ATLAS experiment\footnote{In the ATLAS cell numbering scheme this
corresponds to the centre of the middle compartment cell $\eta_i=27$ (out of $54$) 
and $\phi_i=11$ (out of $16$).} 
of $\eta=0.687$ and $\phi=0.282$.

%Each run typically contained $20000$ events. {\it XXX  to be checked}

The following selection requirements have been applied to select a pure sample of single electrons:
\begin{itemize}
\item the pion counter  had to be %within $280$ and $510$ ADC counts;
      compatible with no signal.
\item the S3 scintillator counter signal had to be % within $750$ and $1500$ ADC counts;
      compatible with that from one minimum ionising particle.
\item cuts on the TDC signals of the chambers
      were imposed to remove double hits and to ensure a good track reconstruction.
      In addition, the beam chamber information was used to define a square of $3$x$3$\cmx$^2$
      around the mean beam position (evaluated for each analysed run) 
      defining the beam acceptance.
\item the $\phi$ and $\eta$ positions reconstructed by the shower barycentre
      must be %$ 10.6 < \overline{\phi} <11.4$ and  $26.6<\overline{\eta}<27.2$. 
      within $0.4$ cell units vertical to the cell centre in $\phi$.
      and within $0.4$ ($0.2$) cell units left (right) from the cell centre.
\end{itemize}

%
%The last requirement ensures that the  $\eta$ and $\phi$ distribution of the impinging beam particles
%(beam profile) is similar in the data and in the Monte Carlo simulation (see section \ref{sec:g4}).

The number of the selected electron events for each energy point can be found 
in Tab.~\ref{tab:beam_momentum}. The run at $E=245$\GeV is left out from the table,
since no precision measurement of the electron beam energy was possible with the
used magnet set-up.
For lower energies the statistics was limited by the rate of electrons in the beam.

%\newpage
\section{Monte Carlo Simulation}
\label{sec:g4}
\noindent 
The simulation of the beam-line and of the calorimeter module
was performed using the \Geant Monte Carlo simulation package \cite{g4}.
The detailed shower development follows all
particles with an interaction range larger than $20$ {\rm $\mu$m}.
Besides purely electromagnetic processes, also hadron interactions, such as those
induced by photon nucleon interactions\footnote{Here, and for the simulations of pions
the QGSP physics list is used for the simulations of hadron interactions.}, 
were simulated.
In addition to the energy deposited in each calorimeter cell, the induced
current  was calculated taking into account the distortion of the electric field in the
accordion structure.
Normalisation factors
equalising the response in the regions of uniform electrical field
were applied to ensure the correct inter-calibration of the accordion
layers.

\begin{figure}[th]
\vspace{-0.5cm}
\begin{center}
\psfig{figure=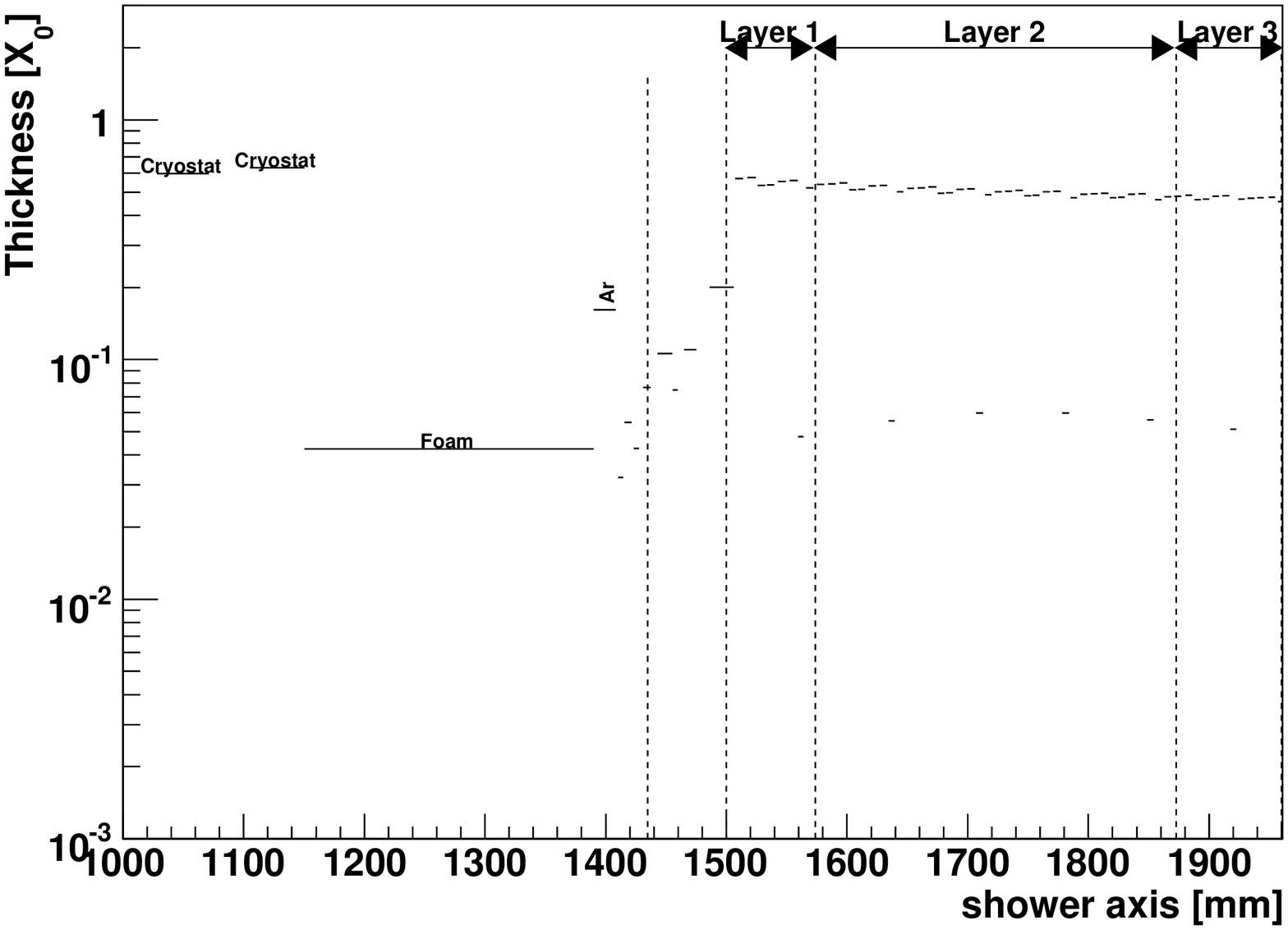,
bbllx=0,bblly=0,bburx=570,bbury=730,angle=270,clip,width=14.cm}
\psfig{figure=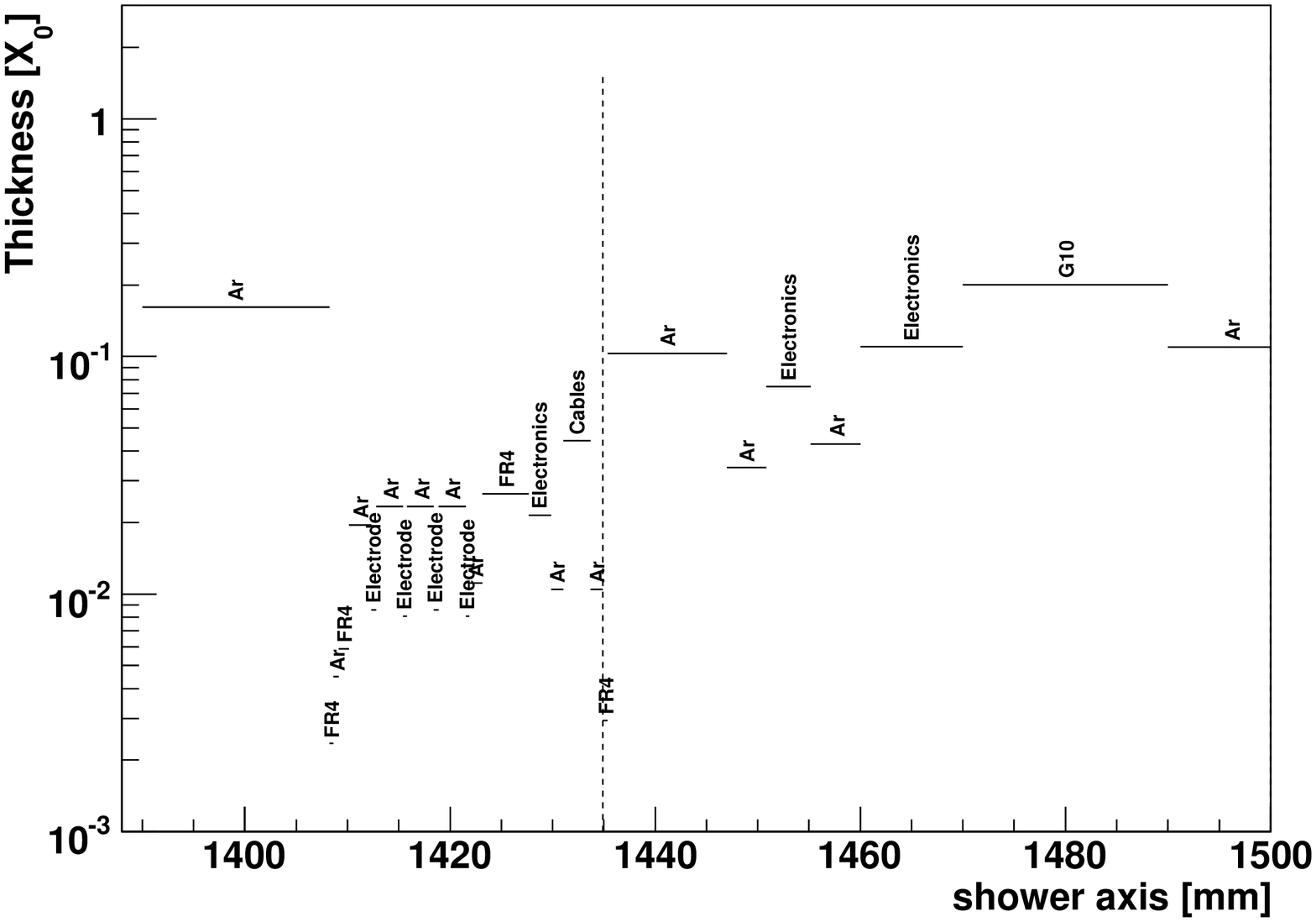,
bbllx=0,bblly=0,bburx=570,bbury=730,angle=270,clip,width=14.cm}
\end{center}
\begin{picture}(0,0) 
\put( 10, 110){a)}
\put( 10,  0){b)}
\end{picture}
\vspace{-0.3cm}
\caption{
 a) Amount of material in the beam line starting just before the cryostat
    along the axis of a particle penetrating at $\eta=0.687$.
 b) Detailed view of the material in the presampler and before the first compartment.
\label{fig:x0vscm}}
\end{figure}

One major challenge in the simulation is the correct description of the passive
material in front of the detector and between the PS and the first accordion compartment.
The details of the material description as implemented in the
Monte Carlo simulation are shown in Fig.~\ref{fig:x0vscm}a and Fig.~\ref{fig:x0vscm}b.
The beam instrumentation before the cryostat corresponds to $0.2$\Xzerox.
The two aluminum walls of the cryostat, the argon excluder (Foam) and the \LAr in front of the
PS have in total a thickness of about $1.5$\Xzerox.

The amount of \LAr between the PS and the inner cryostat wall
is not well known, since the exact position of the argon excluder in front of the calorimeter
was not  precisely measured.
An estimate of $2$\cm  
has been obtained by simulating different configurations and by requiring that the ratio
of the visible energy in the simulation and in the data does not depend on the 
beam energy for each calorimeter layer. 
From this study, a systematic uncertainty of $\pm 0.5$\cm is estimated.

The electronic read-out chain and the signal reconstruction are only partly simulated.
The current to energy conversion takes into account the convolution of the signal
with the shaper response and its integration time. Thus, the response at the peak
of the signal is simulated. 
A cross-talk correction 
derived from calibration runs (see section~\ref{sec:elec_calib})
is applied to simulate the effects on the shape of the energy distribution
in the first compartment. The total energy in the first layer is not modified.

The electronic noise has been extracted from randomly triggered events where the signals
have been reconstructed in the same way as in physics events (see section~\ref{sec:calib}).
This noise has been added incoherently to the energy of each cell.
%Since low energy cells are reconstructed in the high gain configuration
%while high energy cells are reconstructed in medium gain  (see section~\ref{sec:elec_calib}), 
%the noise is slightly lower in low energy cells
%than in high energy ones. 
In the medium gain the noise is slightly larger due to the contribution of the second stage
noise of the electronics.
The noise has been measured in special runs,
where randomly triggered events are recorded with fixed electronic gains
(see section~\ref{sec:pedestal}).
The final noise has then been calculated as the root mean square of 
one of the two samples
chosen according to the probability that a given cell is in high or medium gain.

An event sample was simulated for each energy point in Table.~\ref{tab:beam_momentum}. 
At low (high) energy the simulated event samples were about $20$ ($2$) times larger than the 
corresponding data samples. 
%%% XXXX number to be updated

%%%%%%%%%%%%%%%%%%%%%%%%%%%%%%%%%%%%%%%%%%%%%%%%%%%%%%%%%
%%%%%%%%%%%%%%%%%%%%%%%%%%%%%%%%%%%%%%%%%%%%%%%%%%%%%%%%%%%%%%%%%g

%\newpage
\section{Electronic Calibration}
\label{sec:calib}
\noindent 
The ionisation signal from the calorimeter is brought via cables in the \LAr
out of the cryostat to the front end crates (FEC).
These crates, directly located on the cryostat, house both the
Front End Boards (FEB) and the calibration boards.

On the FEB, the signal is first amplified by a current sensitive preamplifier. In order to accommodate the large
dynamic range and to optimise the total noise (electronics and pile-up), the signal is shaped with a
CR-RC$^2$ architecture (bipolar shape) and split in three linear scales with a typical ratio $1 : 9.2 : 92$, called
low, medium and high gain. For a given channel these three signals 
are sampled at the  $40$ {\rm MHz} clock frequency and 
stored in an analog pipeline (Switched Capacitor Array) until the trigger decision. 
After a trigger, a predefined number of samples ($N$) is digitised by a $12$ bit ADC.
This digitisation is done either on each gain or on the most
suited gain according to a hardware gain selection based on the amplitude of a fixed sample in the
medium gain. 
%As in the beam test, the trigger is not synchronous with the $40$ {\rm MHz} clock, 
%the phase between a trigger and the clock is measured with a TDC.

The response dispersion of the electronics read-out is about $2$\%. 
To account for such an effect 
and for the different detector capacitances  of each calorimeter cell, the calibration
board provides to all channels an exponential signal that mimics the
calorimeter ionisation signal. 
This voltage signal is made by fast switching of
a precise DC current flowing into an inductor and
is brought to the motherboard on the calorimeter via a $50$ $\Omega$ cable terminated at both
ends. The amplitude uniformity dispersion is better than $0.2 \%$.
One calibration signal is distributed through precise resistors
to $8$ ($32$) calorimeter cells for the middle (front and back) 
compartment whose location is chosen such that cross-talk can be studied.

Details on the calibration of the electronics can be found in Ref.~\cite{modul0}. 
%For the $2002$ data taking period a publication
%on the uniformity of the calorimeter response at a fixed energy
%is in preparation \cite{uniformity}. 
%There the full calibration of the electronic chain will be discussed. 
Here we summarise those aspects which are relevant for the linearity of the energy measurement.

The cell energy is reconstructed from the measured cell signal using:
\begin{eqnarray} 
E^{vis}_{cell} =  \frac{1}{f_{I/E}}  \; 
F_{gain}   \; 
\sum_{sample = 1,N} OF_{sample,gain} \; 
(S_{sample} - P_{gain}), 
\label{eq:calib_summary}
\end{eqnarray}
where 
$S_{sample}$ is the signal measured in ADC counts in $N$ time slices,
$P_{gain}$ is the pedestal for each gain (see section~\ref{sec:pedestal})
and $OF_{sample,gain}$ are the optimal filtering (OF) coefficients derived from the shape
of the physics pulse and the noise  (see section~\ref{sec:of}). 
%The OF coefficients are used to
%determine the amplitude and the arrival time of the physics signal. 
The function $F_{gain}$ converts for each gain
ADC counts to currents in $\mu A$ 
%and is measured by injecting a well-known current  in the detector capacity 
(see section~\ref{sec:elec_calib}).
The factor $f_{I/E}$ takes into account the conversion from the measured current 
to the energy (see section~\ref{sec:fieps}).

\subsection{Pedestal Subtraction}
\label{sec:pedestal}
\noindent 
%Due to various electronic effects, e.g. temperature variations, the absolute meaning of a measured ADC count 
%can vary with time. 
In order to determine the signal levels where
%ADC count value where 
no energy is deposited in the detector, special runs with random triggers
and no beam were taken ("pedestal runs"). 
%For the reconstruction of the runs with
%particles impinging on the detector, these pedestals are subtracted.
The stability of the pedestal values was checked using runs taken in regular
intervals throughout the data taking period. A run-by-run instability 
was  observed in particular for the PS
which has a non-negligible effect on the
reconstructed energy. No instability has been observed within a run.
To minimize the effects of such instabilities each electron run was corrected using
pedestal values measured with random triggers within the same run.
%In order to avoid these instabilities
%the pedestals have been recalculated using random triggers taken in parallel with the
%electron events. 
This ensured that for each physics run the correct pedestals are calculated and
possible biases  of about $20$\MeV are corrected.
%which would cause a non-linearity of the reconstructed electron energy of up to $0.2-0.3$\%.

\subsection{Determination of the Signal Amplitude}
\label{sec:of}
\noindent 
%The signal induced by the ionisation of the \LAr in a calorimeter cell is 
%amplified, shaped and sampled
%at the LHC bunch crossing frequency every $25$\ns and is digitised.
%
The peak amplitude $A$ (and the signal time) is extracted from the $N=5$ signal
samples ($S_{sample}$)
using a digital filtering technique \cite{of}. 
The peak amplitude is expanded in a linear weighted sum
of coefficients (OF) and the pedestal subtracted signal in each sample
(see eq.~\ref{eq:calib_summary}).

The coefficients are calculated  using the expected shape of the physics signal,
its derivative and the noise autocorrelation function. %in all samples.
%The calculation is based on a Lagrange multiplier technique where 
The noise contribution is minimised respecting constraints on the signal
amplitude and its time jitter.
The noise autocorrelation function is determined from randomly triggered events.

%The physics pulse shape is determined in a semi-predictive approach \cite{of2,unitarity}:
%The method exploits that most of the signal path is the same for the physics pulse
%from the ionisation in a cell and the pulse injected by the calibration system\footnote{The full
%read-out chain from the motherboard to the front-end board is common.}.
%The calibration pulse shape is reconstructed for each cell in special calibration runs
%(delay runs), where a signal with a fixed amplitude and a variable time for the pulse is injected.
%Each calibration pulse is composed out of $175$ samples with a $1$\ns sampling step.

%To predict the physics pulse shape from this signal two points have to 
%be taken into account: First, the physics pulse stems from a triangular input signal 
%while the calibration signal stems from an exponential one. 
%Second,  the calibration signal is injected on the motherboard and is not generated
%in the \LAr cell. The path between the \LAr cell (modelled with a capacitance $C$)
%and the motherboard can be modelled by an inductance ($L$) and a resistor ($r$).
%Using a simple electrical model of a read-out channel, the form of the physics channel
%can be predicted using a formula
%with four free parameters: $\omega_0=1/\sqrt{C L}$, $\tau = r C$
%and the starting times of the calibration and the physics pulse. 
%The free parameters can be extracted from a fit to the
%measured physics pulse shape. For each cell the OF coefficients are then calculated \cite{prieur}.

The shape of the physics signal can be predicted using a formula with four free parameters
that can be extracted from a fit to the measured physics pulse shape 
\cite{of2}. %\cite{of2,uniformity}. 
For each cell, the OF coefficients are then calculated \cite{prieur}.
Over the full module, the shapes of the measured and predicted physics pulse 
agree to within $2\%$ and the residuals 
of the pulse shape at the peak position are within $0.5$\% in the first and $1$\% in the second compartment.
For the cells involved in the electron energy measured at the beam position studied in this analysis,
the residuals deviate by at most $7 \cdot 10^{-3}$ in the first compartment and
by at most  $2.5 \cdot 10^{-3}$ in the second compartment.
%The normalisation difference between the physics and the
%calibration signal is deduced and applied.

Due to the fine segmentation of the first calorimeter compartment
there is unavoidably a capacitive coupling between 
the read-out cells (strips). 
This cross-talk affects the signal reconstruction during 
the calibration procedure as well as the physics pulse shapes.
The cross-talk (see section~\ref{sec:elec_calib})
has been taken into account in the determination of the OF coefficients.

\subsection{Calibration of the Read-out Electronics}
\label{sec:elec_calib}
\noindent 
The relation between the current in  a \LAr cell ($\mu A$)
and the signal measured with the read-out electronics (in  {\rm ADC} counts) is determined by injecting
with the calibration system a well known current. 
%To account for the variations of the electronics with time, 
Calibration runs have been taken in regular intervals (about every $12$ hours) for each cell.
In these calibration runs currents with linearly rising amplitudes (in  {\rm DAC} units) 
are injected ("ramp runs").

%To ensure a dynamic range from a few\MeV to several\TeV 
%with a signal of a fixed bit size, three different amplification factors (gain) are used.
%In the test-beam, where only an energy range from $10\GeV$to $245\GeV$was available,
%only high and medium gain are needed.
%Typically a cell is read-out in the medium gain, if the energy in this cell is 
%about above 23\GeV in the second compartment.

For each cell and for each injected current, the amplitude is reconstructed from the measured
signal (after subtracting a parasitic injected current) % at low DAC values) 
adjusting the pulse shape derived from special calibration runs (``delay runs''), 
where a signal with a fixed amplitude and a variable time for the pulse is injected.

Each reconstructed amplitude rises almost linearly with the
input signal. To correct for small non-linearities the dependence of the reconstructed amplitude
on the input signal is fitted with a fourth order polynomial. 
This fit is used to reconstruct the visible energy in a \LAr cell 
($F_{gain}$ in eq.~\ref{eq:calib_summary}).

\begin{figure}[th]
\vspace{-0.5cm}
\begin{center}
\psfig{figure=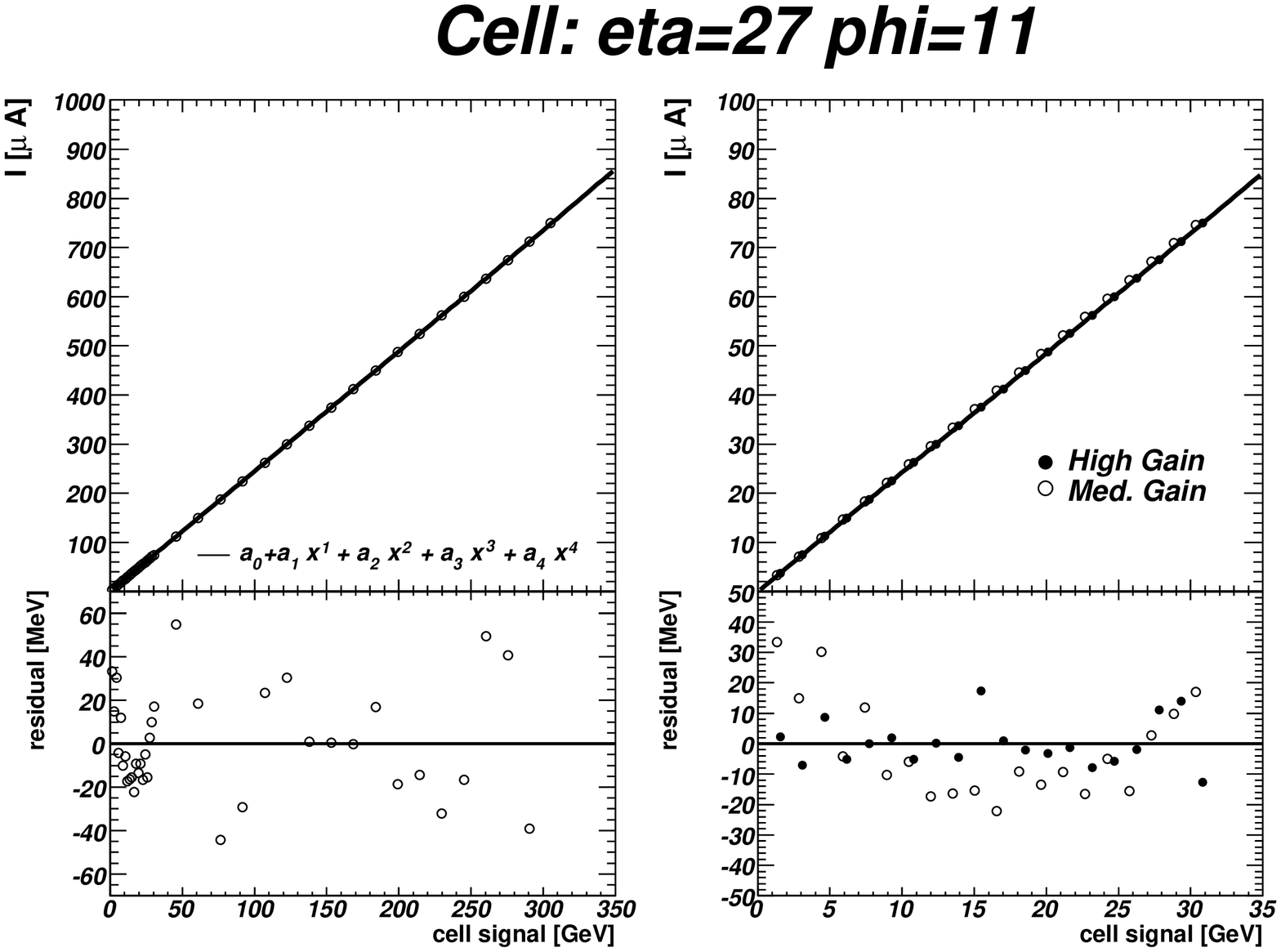,
bbllx=109,bblly=57,bburx=570,bbury=730,angle=270,clip,width=15.cm}
\end{center}
\begin{picture}(0,0) 
\put(  5, 5){a)}
\put( 75, 5){b)}
\end{picture}
\vspace{-0.3cm}
\caption
{Relation of the measured cell signal in\GeV and the current
 ($\mu A$) injected by the calibration system for high and medium
 gain. 
 Superimposed is the result of a fit to a fourth order polynomial. 
 %In the top insert the
 %deviation from linearity, i.e., the difference of the injected current and the
 %linear part of the fit divided by the injected current,
 %is shown. 
 In the bottom the residual, the difference of the injected current
 and the fit, is shown as a function of the measured signal.
 In a) the results for the medium gain and in b) the results for the high
 gain are shown. In addition, in b) the low signals of the medium gain are superimposed.
\label{fig:ccalib}}
\end{figure}

As an example, in Fig.~\ref{fig:ccalib} 
the results of the ramp run analysis is shown 
for the cell that %has for most events
had on average the largest energy fraction during the data taking period.
The relation between the injected current in $\mu A$ and the reconstructed
amplitude in\GeV is shown. For a given DAC value on the calibration board a current
is injected and the resulting signal is measured in ADC counts.
To facilitate the interpretation\footnote{
%In the middle sampling $1$ DAC count in the
%calibration board corresponds to $37.5${\rm nA} ($f_{DAC/nA}$). 
%The conversion of the measured ADC counts to\GeV has been approximated by:
%$a_1 f_{DAC/nA} / d (f_{I/E} f_{samp})$, where
%$a_1$ is the first coefficient of the polynom $P_4$ 
%($a_1=7.1$ ($0.76$) {\rm DAC/ADC} for medium (high) gain),
%$f_{I/E}=16$ {\rm nA/MeV} (see section~\ref{sec:fieps}),
%at $E=100$\GeV 
%$f_{samp}=0.18$ (see eq.~\ref{eq:ErecAcc}) at $d=0.854$ and
%is a correction for reconstruction effects.
To convert ADC counts to GeV $f_{ADC/GeV}= 0.1082 \; (0.0117)$ has been used
for  medium (high) gain.} 
the data are already transformed into in\GeV units.

The data obtained in medium gain are shown as open circles. 
Data in high gain are shown as closed circles. At low energies, where the signal can be reconstructed in both gains
good agreement is found. In the region where the gains are switched the
difference between the high gain and the medium gain is only a few\MeVx. 
The deviation from linearity of the electronics for a cell signal is $1-2 \%$.
The line shows the result of a fit of a fourth order polynomial $P_4({\rm ADC})$.
%The insert illustrates the deviation from linearity\footnote{The deviation from linearity
%is defined as $100 \, ({\rm DAC } -  (a_{0} +  a_{1} {\rm ADC}))/{\rm DAC}$.}.
The bottom part of the figure shows the residuals, i.e., 
the difference between the reconstructed and the input
signals in\MeVx. For the medium (high) gain the signal is reconstructed with an accuracy
of $40$ ($10$)\MeVx.

%%%%%%%%%%%%%%%%%%%%%%%%%%%%%%%%%%%%%%%%%%%%%%%%%%%%%%%%%
%%%%%%%%%%%%%%%%%%%%%%%%%%%%%%%%%%%%%%%%%%%%%%%%%%%%%%%%%%%%%%%%%g

During the calibration of the electronics the cells are pulsed with 
a pattern where between two pulsed cells three cells are not pulsed.
Since the calibration constants are derived from the known injected current and
the measured signal for the pulsed cells, they are overestimated if the
effect of cross-talk is not taken into account.
In the first compartment, where the capacitive coupling between the cells is large,
a cross-talk correction is derived for each cell using delay runs. It is obtained
by adding the signal measured in
the two closest passive neighbours of pulsed cell and the average of
the next-to-closest neighbour. 
%The average can be used, since these cells
%receive a similar contribution from the next pulsed cells 
%as that from the cell under investigation. 
%The fitted peak-to-peak ratio
%between the pulse shape with and without cross-talk is the correction
%to the calibration constants. 
The correction factors
do not strongly vary for different electronic gains and they are moreover 
stable for all pulse values.

\subsection{The Current to Energy Conversion Factor}
\label{sec:fieps}
\noindent 

The conversion from
the measured current ($\mu {\rm A}$) to the visible energy (\MeVx) 
is done using a factor $f_{I/E}$, which is assumed to be independent of the beam energy. 
There is one common factor for the three accordion compartments and one factor for the
presampler. Both factors are difficult to calculate from first principles.
The difficulties arise from the complex structure of the
electric field and from the modelling of physical effects like recombination
of electrons in the \LArx. 

In this analysis they are determined
from a comparison of the visible energies in the data and in the Monte Carlo simulation,
where the complex accordion geometry and in particular the
electrical field are simulated in detail. In this way also the dependence of the simulated
signal on the range cut, below which a particle is not tracked any further and
deposits all its energy, is absorbed.
Since the current to energy conversion drops out in the linearity analysis, its exact
value is not  critical.

The $f_{I/E}$ factor can be roughly estimated using a simplified model,
where a detector cell is seen as a capacitor
with constant electrical field:
$f_{I/E}= e/(w * t) \approx 15$ {\rm nA/MeV}, 
where $e$ is the elementary charge, $w$ is the ionisation
potential of \LAr and $t$ is the drift time.
The comparison of the data to the Monte Carlo simulation gives in the accordion calorimeter  
$f_{I/E} =  16.0$ {\rm nA/MeV}. 

%Due to different cable lengths in the calibration system, the
%resistances for the first and the second compartment (see section~\ref{sec:elec_calib})
%are slightly different than the nominal values assumed in the reconstruction software.
%
%The signal reconstructed
%in the first compartment is therefore attenuated by $0.993$ with respect to that 
%of the second compartment.
The calibration signal amplitude is attenuated due to the skin effect
in the cables. The different lengths used in each compartment result
in a small bias and the front signal has to be corrected by a factor $1.007$.
Differences from the slightly different electrical fields in the first and the second
compartment (due to the different bending of the accordion folds) are estimated
to be of the order of $0.3\%$ (using calculations of the electric field). 
The cross-talk effect in the first compartment (see section~\ref{sec:elec_calib})
is corrected cell-by-cell. The result of this correction is that 
on average the measured total signal is lowered by
a factor $0.93$. For all these effects,
an uncertainty of  $\pm 0.5\%$ is assigned for the
relative normalisation of the  first and the second compartment.

In addition, a correction for cross-talk between the second and the third compartment is also needed. 
This is due to the read-out lines of the second compartment passing through the third one.
Empirically it has been found that
$0.55\%$ of the energy deposited in the second compartment ($E_2$) is measured in the
third compartment ($E_3$). Therefore, $0.55\%$ of $E_2$ is subtracted from $E_3$ and added
to $E_2$. The overall energy is not changed by this correction. However, to
compare the energy fraction deposited in the individual layers and the
mean shower depth in data and Monte Carlo simulations it has to be taken
into account.

In the PS the current to energy conversion 
factor is also estimated to be about $f_{I/E} =  16$ {\rm nA/MeV}.  
Two effects must be taken into account that reduce this factor:
First, one cell with coherent noise had to be excluded from the analysis.
For the impact point studied here this 
leads to a reduction of $0.95$ of the total PS signal (according to the Monte Carlo simulation).
Second, the effective length of the PS is reduced to $11$\mm 
due to the vanishing electric field at  
the edges, the factor is  $f_{I/E} =  16 \cdot (11/13) \cdot 0.95 = 12.9$ {\rm nA/MeV}.
This is in agreement with the number found from the comparison of data to Monte Carlo.

%\newpage
\section{Calibration of Electromagnetic Showers}
\label{sec:calibration_emshower}
\subsection{Calibration Constants for Sampling Calorimeters}
\label{sec:calibration_long_comp}
\noindent 
In a sampling calorimeter the total deposited EM energy ($E^{tot}$) 
can be estimated from the energy deposited in the active medium ($E^{act}$)
by dividing by the sampling fraction \fsamplex:
\begin{equation}
E^{tot} = \frac{1}{\fsample} \; E^{act}, \;\; {\rm with} \;  \; \fsample 
= \frac{E^{act}}{E^{act}+E^{pas}},
\label{eq:sampling}
\end{equation}
where $E^{pas}$ is the energy deposited in the passive material.

For a minimum ionising particle the sampling fraction is a fixed number which can
be calculated from the known energy deposits 
in the active and passive materials due to ionisation.
Since the energy loss of electrons is different from that of muons, the sampling fraction
for electron is lower.
%
%in addition energy by radiation, the $e/\mu$-ratio 
%,which can be obtained from empirical formulae using masses of the materials in the calorimeter, 
%has to be taken into account.

\begin{figure}[th]
\begin{center}
\psfig{figure=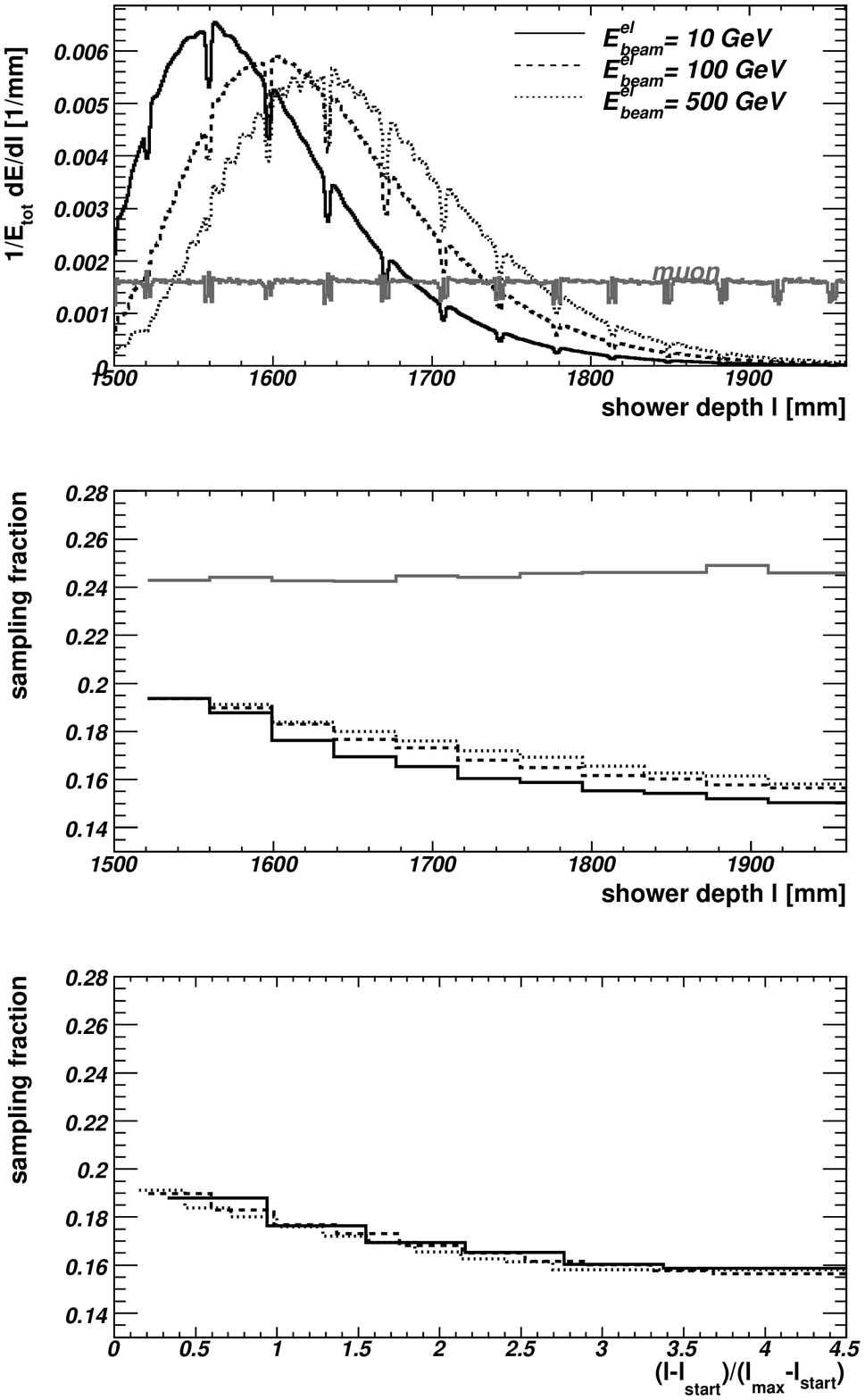,
bbllx=40,bblly=40,bburx=530,bbury=760,clip, %angle=270,clip,
width=14.cm}
\end{center}
\begin{picture}(0,0) 
\put(  5,145){a)}
\put(  5,75){b)}
\put(  5, 5){c)} 
\end{picture}
%\vspace{-0.5cm}
\caption{
a) Shape of the energy deposited along the shower axis.
b) Sampling fraction along the shower axis.
c) Sampling fraction rescaled to the shower maximum.
Shown are Monte Carlo simulations for electrons with
beam energies of $10$, $100$\GeV and $500$\GeV and muons with  $10$\GeVx.
In c) $l_{start}$ is defined as the beginning of the calorimeter
and  $l_{max}$ is the maximum of the shower. 
\label{fig:cprof_rebin}}
\end{figure}

In Fig.~\ref{fig:cprof_rebin}a is shown the shape of the deposited
energy distribution of an EM shower
along the shower axis ($l$) for electrons with $E= 10$\GeVx, $E= 100$\GeV and $E= 500$\GeVx,
and for muons with  $E= 10$\GeVx. 
The accordion calorimeter starts at $1500$\mmx. 
The sampling structure of the calorimeter ends at about $1960$\mmx. 
%After that the particles loose their energy in the \LAr of the cryostat.
The energy depositions before and after the calorimeter are not shown. 

The muon loses only a small part of its energy in 
the calorimeter. The dips approximately every $40$\mm are 
due to the lead traversed by the muon going through the accordion (zig-zag) folds.
%caused by the changing passive material due to the absorber folds.

While the energy deposited by muons
is approximately constant, for EM showers
the deposited energy rises quickly to a maximum and then is slowly attenuated.
As the energy of the impinging particle increases, the shower penetrates deeper
into the calorimeter. In most events the shower is contained inside the calorimeter
and only a very small fraction of the energy leaks out. 
At the end of the shower more and more particles at low energy are produced.

Particles produced in the EM shower interact differently
with the detector at the beginning and at the end of the shower development.
%At the end of the shower a large number of low-energetic photons are produced which
%have a higher probability to be stopped in the lead absorbers than in the \LAr.
%
At the end of the shower a large number of low-energetic photons are produced,
which have a higher probablity to produce 
low energy electrons (via e.g. the photo-electric effect and Compton scattering)
in the lead absorber than in the Lar. Since the range of these electrons is typically 
smaller than the lead absorber thickness, the energy
deposit in the absorber increases relative to the energy deposit in the active material
towards the end of the shower and the sampling
fraction decreases \cite{gg,pinkau,beck,Crannell:1969ag,Flauger,delPeso:1990er}. 
This is illustrated in Fig.~\ref{fig:cprof_rebin}b, where the sampling fraction along the shower
axis is  shown for the same particles as in Fig.~\ref{fig:cprof_rebin}a.
While the sampling fraction is constant for muons,
for electrons the sampling fraction decreases towards the end of the shower.
For an electron with $E= 10$\GeV the sampling fraction drops by $20$\%.
This behaviour depends on the electron energy.
However, when the sampling fraction is calculated as a function of the relative distance
from the shower maximum, it shows a universal behaviour, i.e., it does not depend on the
electron energy \cite{gg} (see Fig.~\ref{fig:cprof_rebin}c).

When looking at a fixed point of the calorimeter, the sampling fraction
can be different event-by-event due to longitudinal shower fluctuations.
% the shower will start earlier or later in the calorimeter and therefore 
%the sampling fraction 
%at a fixed point of the calorimeter.
If, however, the response of the calorimeter is equal in all regions, 
%the energy is deposited in the same way over the whole calorimeter, i.e., 
the sampling fraction is the same %for all beam energies 
when integrated over the whole shower depth. 
%$\fsample = \int_0^\infty \fsamplex(l) \;  dl$. 
Therefore the calorimeter is linear.
Nevertheless, a non-linearity is introduced, if only a part
of the shower is contained in the calorimeter or if the longitudinal compartments are not
equally calibrated, i.e., if they react differently to minimum  ionising particles. 
%Even so, for a given shower, in principle each longitudinal compartment has a different sampling fraction, all
%compartments have to be calibrated in the same way. Otherwise, a non-linearity is introduced
%due to the longitudinal fluctuations of the EM shower and due to the dependence of its penetration
%depth on the beam energy. 
%
In most practical applications, the shower starts already upstream of the calorimeter and a fraction of its
energy is also deposited behind the calorimeter. This introduces an intrinsic non-linearity of the
energy response due to longitudinal shower fluctuations,
which must be corrected.

%%%%%%%%%%%%%%%%%%%%%%%%%%%%%%%%%%%%%%%%%%%%%%%%%%%%%%%%%
%%%%%%%%%%%%%%%%%%%%%%%%%%%%%%%%%%%%%%%%%%%%%%%%%%%%%%%%%%%%%%%%%%%%%

\subsection{Correction for Upstream Energy Losses using Presampler Detectors}
\label{sec:upstream_correction}
\noindent 
The principle of a presampler detector is that the energy deposited in a thin active medium
is proportional to the energy lost in the passive medium in front of the calorimeter.
The difference with respect to a sampling calorimeter is that the
passive material, the ``absorber'', is very thick, typically about $1-2$ \Xzerox, and that
there is only one layer of passive and active material.
Therefore, the calibration scheme is different from the one of
sampling calorimeters where a shower passing a radiation length
is sampled many times. % like the one of eq.~\ref{eq:sampling}.

If an electron passes through the passive material in front of the presampler,
it continuously loses energy by ionisation.
The total deposited energy in the passive material is approximately constant and can be calculated
assuming the energy is lost by a minimally ionising particle.
However, the electron also emits photons by Bremsstrahlung.
Depending on their energy they either react through Compton scattering or photo-electric effect
(in this case their energy is mainly deposited in the dead material by low energy electrons) 
or they do not interact until they create an electron positron ($e^+ e^-$) pair. 
The pair can be produced in the passive material, in the
active medium of the presampler or in the sampling calorimeter. In the latter case, 
their energy is simply measured in the
calorimeter and nothing has to be done in addition.
%In the first case, the pair will be produced towards the end of the passive material,
%since the probability to create a $e^+ e^-$ pair is proportional to the exponential of the photon path length.
%In the first case, each of the electrons produced by the photon deposits less energy than the one deposited
%by the beam electron passing through the full passive material.
The energy deposited by each particle of
the  $e^+ e^-$ pair is therefore in many cases smaller than the energy deposited
by the beam electron passing through the full passive material. 
%In the presampler, however,
%the  $e^+ e^-$ deposits approximately twice the energy of the beam electron. 
In the case where the pair is created
in the presampler itself, the energy measured in the presampler is even largely uncorrelated to the
energy deposited in the passive material.

In any case the  electrons produced by pair-production will traverse none or part of 
the passive material. If one  $e^+ e^-$-pair
is created in the passive material, three electrons ionise the active medium.
Two of them have only traversed a small part of the passive material. 
The correct calibration constant
is therefore smaller than the one calculated from the inverse sampling fraction of a 
minimum ionising particles.

The total energy deposited in the active and in the  passive material 
in and before the presampler  $E^{tot}_{0}$
can be reconstructed from:
\begin{eqnarray}
E^{tot}_{0} \; \; = w_{PS} \; E^{act}_{0} \; \; = 
%\frac{ E^{act}_{0} + E^{pas}_{0}}{E^{act}_{0}} \;  E^{act}_{0} \; \;
 \frac{a + b \; E^{act}_{0} }{E^{act}_{0}} \; \;  
E^{act}_{0} = a + b \; E^{act}_{0}
\end{eqnarray}
where %$E^{tot}_{0}$ is the reconstructed energy deposited in and before the PS and
$E^{act}_{0}$ is the energy deposited in the active medium of the presampler. % and 
%$E^{pas}_{0}$ is the energy deposited in the passive material in and before the presampler.
The calibration coefficient
$a$ represents the average energy lost by ionisation by the beam electron.
Its energy dependence might be caused by low energy photons produced by Bremsstrahlung
that are absorbed in the dead material and by photon nucleon interactions.
The amplification factor $b$ takes into account
that the  $e^+ e^-$-pairs produced in the passive material or
in the active medium have only traversed part or none of the 
material in front of or in the PS. 

The calibration factors depend on the details of the experimental set-up and have
to be extracted from a Monte Carlo simulation.

%A similar calibration scheme has already been proposed in Ref.~\cite{gu}. 

\section{Electron Energy Reconstruction}
\label{sec:reco}
\subsection{Reconstruction of the Electron Cluster Energy}
\label{sec:clustering}
\noindent 
When an electron penetrates the ATLAS calorimeter a compact  EM shower is
developed, which deposits most of the energy near the shower axis.
%To reconstruct the electron energy, the energies deposited in several calorimeter
%cells have to be added together. The energy deposits far away from the shower axis get
%increasingly small and they are at some point comparable to the electronic
%noise. Since electronic noise has the same probability 
%to give positive and negative contributions
%the average energy in a cell is not changed by noise. However, to ensure a good energy
%resolution it is better to restrict the electron fiducial envelop to the
%core, i.e. the cells far ways from the shower axis are not used
%to reconstruct the electron energy. 
%The collection of cells from which the electron energy is measured is called ``cluster''.
%
%Since the radial EM shower energy profile does merely depend on the electron
%energy, the signal loss can be easily corrected. This has the additional advantage
%that no noise cut has to be introduced, which would
%inevitably cut out some part of the signal. The introduction of a noise
%threshold would make an energy dependent correction necessary.
%
%
To reconstruct the electron energy, the energies deposited in a fixed number of calorimeter
cells are added together. No noise cut is applied.
This collection of cells is called ``cluster''.
Since the radial EM shower energy profile depends merely on the electron
energy, the signal loss from cells outside the cluster can be easily corrected. 

The electron cluster is constructed from the second accordion compartment,
where all cells within a square of \standclus~cells around the cell with the
highest energy are merged. For the other accordion compartments
all cells which intersect the geometrical projection of this square are included.
In the first accordion compartment, which has high granularity in the $\eta$-direction,
$8$ cells from each side of the cell with the maximum energy in this compartment
are added to the cluster. For the impact point analysed here, two cells in $\phi$ are included. 
Thus, a cluster includes $17{\rm x}2$ cells in the first,
$3{\rm x}3$ cells in the second and $2{\rm x}3$ in the third compartment.
while in the PS $3{\rm x}2$ cells are used to define the visible energy.

To reconstruct the electron energy deposited in the accordion calorimeter
the visible energy measured within the
electron cluster defined above is multiplied by a calibration factor:
\begin{equation}
E^{rec}_{acc} =   \sum_{i=1,3 } E^{rec}_i =  
\frac{1}{d(E) \, \scriptsize f_{samp}}\; \sum_{i=1,3 } E^{vis}_i, 
\label{eq:ErecAcc}
\end{equation}
where $E^{vis}_{i}$ is the visible energy in 
the $ith$ compartment and
$f_{samp}=0.18$ is the sampling fraction taken for an electron with an
energy of $100$\GeV
(see eq.~\ref{eq:sampling}).
%The true sampling fraction in the accordion calorimeter\footnote{This number can be calculated
%from first principles using the accordion calorimeter cell geometry and the electron muon ratio.}
%is $0.18$ (see eq.~\ref{eq:sampling}).

\begin{figure}[th]
\begin{center}
\psfig{figure=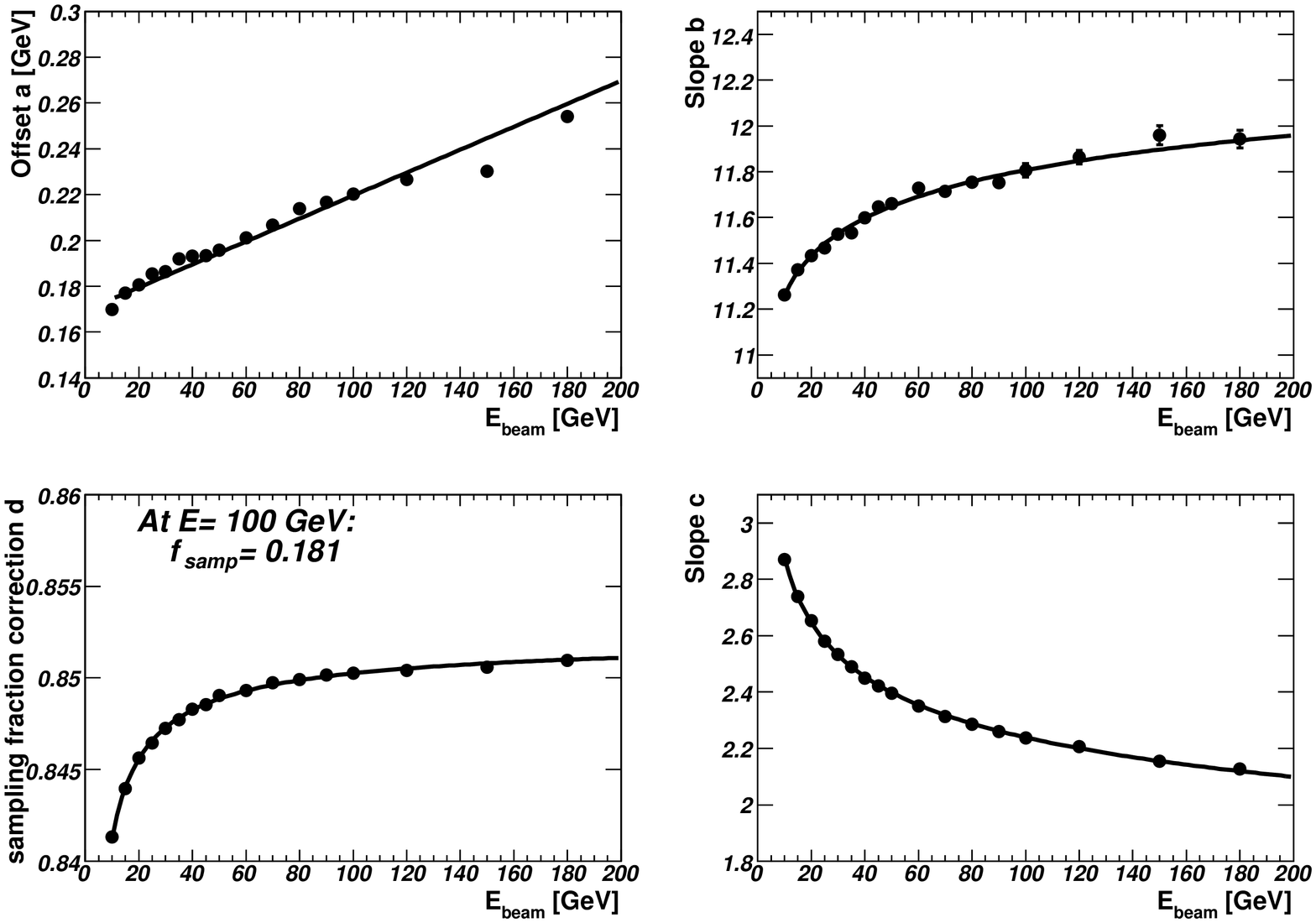,
bbllx=60,bblly=0,bburx=650,bbury=850,angle=270
,clip, width=14.5cm}
\end{center}
\begin{picture}(0,0)
\put(5,   60){a)}
\put(78,  60){b)} 
\put(5,   10){c)} 
\put(78,  10){d)}
\end{picture}
\vspace{-0.5cm}
\caption{Calibration parameters extracted from the 
Monte Carlo simulation as a function of the beam energy.
The calibration parameters are defined in eq.~\ref{eq:Erec}.
The lines illustrate a parameterisation of the energy dependence of the calibration coefficients.
\label{fig:calib_para}}
\end{figure}

The factor $d$ is depends on the initial electron energy
(see Fig.~\ref{fig:calib_para}c). 
Its variation is mainly (about $0.6$\%)
due to the decrease of the sampling fraction towards the end of the shower, and due to 
the fact that the shower has already started before entering the accordion calorimeter
as discussed in section~\ref{sec:calibration_long_comp}. 
In addition, the factor $d$
corrects for a drop of the total energy of about  $10$\% due to 
the  reduced charge collection near the accordion folds %and the HV resistors
and for energy lost laterally (typically  $4$\% of the total electron energy).
The fraction of energy outside the electron cluster varies %% XXX increases or decreases 
by about $0.3$\% with energy. 
The current to energy conversion leads to a decrease of the reconstructed energy
by $0.5$\% at high energies.
Since these effects are correlated, they
do not factorise and are therefore absorbed into one factor.

%%%%%%%%%%%%%%%%%%%%%%%%%%%%%%%%%%%%%%%%%%%%%%%%%%%%%%%%%
%%%%%%%%%%%%%%%%%%%%%%%%%%%%%%%%%%%%%%%%%%%%%%%%%%%%%%%%%%%%%%%%%g

\subsection{Correction for Upstream Energy Losses}
\label{sec:upstreamlosses}
\noindent 
Following the correction procedure for energy losses in the upstream material,
%using the presampler detector has been 
described in section~\ref{sec:upstream_correction},
the total energy deposited in and before the PS is reconstructed by:
\begin{equation}
E^{rec}_{0} = a + b \; E^{vis}_0, 
\label{eq:ErecPs}
\end{equation}
where $E^{vis}_0$ is the visible energy in the PS cluster.
 
The calibration parameters $a$ and $b$ are obtained from the Monte Carlo
simulation. 
As an example, the correlation between the visible energy deposited
in the PS and the true energy deposited upstream and in the PS
is shown in Fig.~\ref{fig:dmcalib}a and Fig.~\ref{fig:dmcalib}b
 for electrons with $E = 10$ and $E = 180$\GeVx.
The calibration parameters $a$ and $b$ are determined by a linear fit.

The offset $a$ rises linearly with the beam energy 
(see Fig.~\ref{fig:calib_para}a) and can be easily parameterised.
The slope $b$ rises logarithmically (see Fig.~\ref{fig:calib_para}b)).
%is almost constant (between $11.2-11.5$), but for a precise
%reconstruction its energy dependence needs to be taken into account.
The value of $b$ corresponds to about $60\%$ of the inverse sampling fraction of a minimally ionising particle
completely passing through the full active and passive medium.

\subsection{Correction for Energy Losses between the Presampler and the Accordion}
\label{sec:psstriplosses}
\noindent 
The region between the PS and the first accordion compartment 
contains support structures,
electronics and cables. The amount of passive material depends on 
the impact point in  $\eta$ and $\phi$.

At this point, a muon with an energy of $10$\GeV deposits 
$0.6$\% of its energy in the passive material in front of the calorimeter
and about $0.15$\% in the passive material between the PS and the first
accordion compartment. 
%This reflects that there is in total more passive material in
%front of the presampler (mainly the cryostat) than there is material between
%the presampler and the first accordion compartment.
%
However, electrons deposit a larger fraction of their energy in the material between the PS and 
the first accordion compartment. % is more significant, since there are more particles produced
%by the EM shower passing through this region. 
An electron with $10$\GeV deposits on average
$3.6$\% of its energy in the passive material in front of the calorimeter and
$4.1$\% in the passive material between the PS and the first
accordion compartment. 
The fact that more energy is deposited behind the PS than before
%This means that more energy is lost after the PS than before. This effect 
gets more pronounced towards higher energies.
An electron with $180$\GeV deposits
$0.45$\% of its energy in the passive material in front of the calorimeter and
$0.85$\% in the passive material between the PS and the first
accordion compartment. 

The energies deposited before and just after the PS 
are linked via the dynamical behaviour
of the EM shower development. 
%Since this correlation is rather complex, it is better to
%correct with the measured PS energy, the energy deposited before the
%PS and with a different observable the energy between the
%PS and the  first accordion compartment.
%In this way, one can correct better effect-by-effect.
%
According to the Monte Carlo simulation, a good correlation
to the energy deposited between the PS
and the first accordion compartment can be obtained from an observable combining the
energy in the PS and the energy measured in the first accordion
compartment, namely:
\begin{eqnarray}
E^{rec}_{PS/Strip} = c \; {(E^{vis}_{0} \cdot E^{vis}_{1})}^{0.5}.
\label{eq:EpsStrip_rec}
\end{eqnarray}
The value of the exponent in eq.~\ref{eq:EpsStrip_rec} has been found empirically.
%It is connected to the dynamics of an EM shower, but can not be easily calculated from first principles.
The calibration coefficient $c$ is obtained from the Monte Carlo simulation. 
The result of the simulation is shown in Fig.~\ref{fig:calib_para}d) as a function of the
beam energy.
%As shown in Fig.~\ref{fig:calib_para}d,
%the parameter $c$ is around $2.8$ and decreases toward $2$ for higher energies.

\begin{figure}[th]
\begin{center}
\psfig{figure=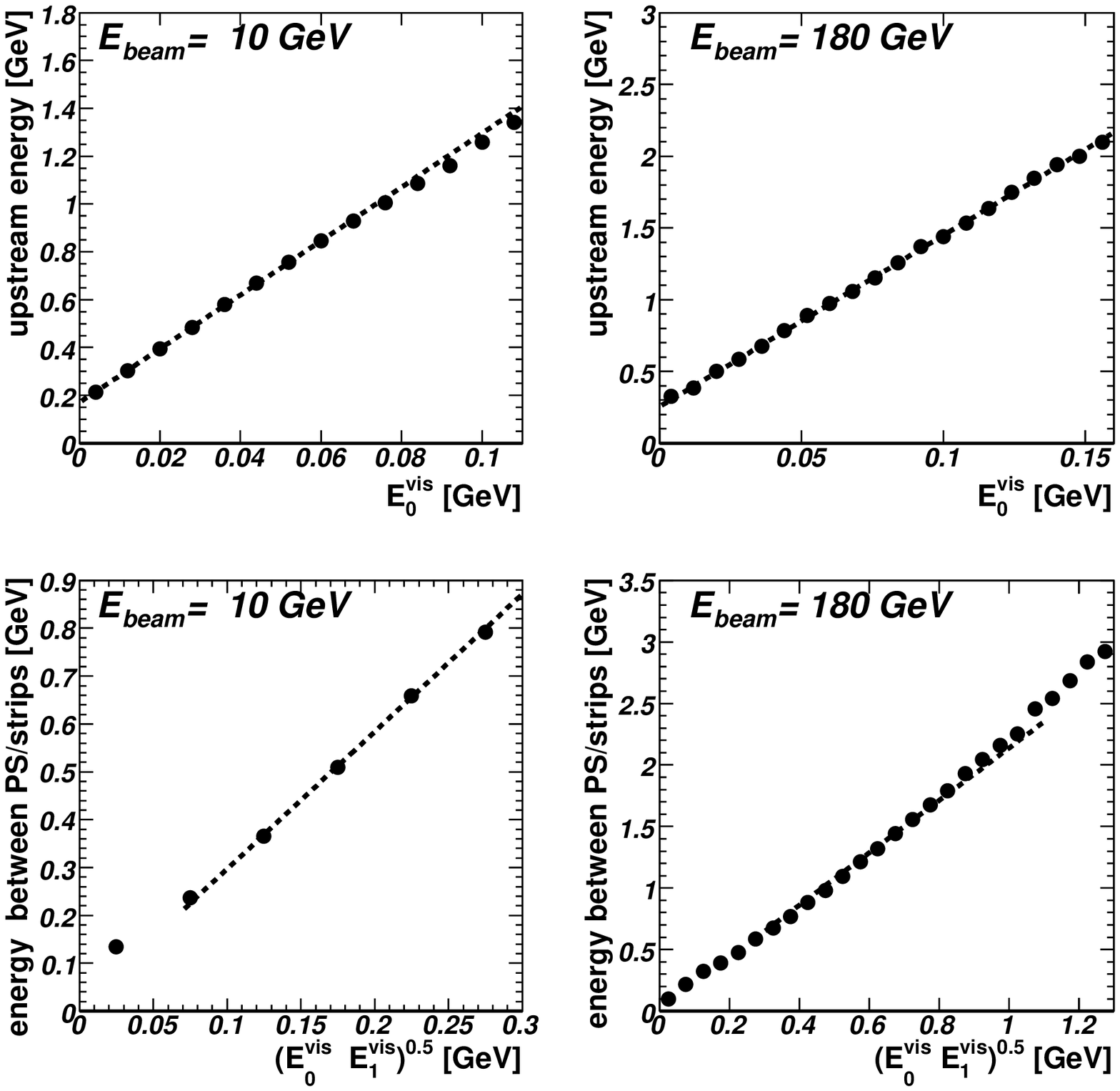,
%bbllx=0,bblly=0,bburx=560,bbury=630,angle=270,clip,
width=14.cm}
\end{center}
\begin{picture}(0,0) 
\put(  5, 5){c)} 
\put( 75, 5){d)}
\put(  5,80){a)}
\put( 75,80){b)}
\end{picture}
\vspace{-0.5cm}
\caption{Mean energy lost before and in the PS as a function of the visible energy in the PS
         for electrons of $10$ (a) and $180$\GeV (b).
         Mean energy lost after the PS and before the first accordion compartment
         as a function of the estimator $(E^{vis}_{0} \cdot E^{vis}_{1})^{0.5}$
         for electrons of $10$ (c) and $180$\GeV (d).
         The dashed lines indicate the linear approximation in the range, where the calibration
         parameters have been extracted.
\label{fig:dmcalib}}
\end{figure}

As an example, the correlation between ${(E^{vis}_{0} \cdot E^{vis}_{1})}^{0.5}$
and the true energy deposited between the PS and the first compartment
is shown in Fig.~\ref{fig:dmcalib}c and  Fig.~\ref{fig:dmcalib}d
for electrons with $E = 10$ and $E = 180$\GeVx.
The calibration parameter $c$ is obtained as the slope of a linear fit.
While this assumption is justified for low energies, at
high energies deviations from a linear correlation are observed.
The exponent in eq.~\ref{eq:EpsStrip_rec} seems to be slightly energy dependent.
Since, however, the dead material correction is less important at high energies, 
within the present accuracy this effect can be neglected.
%

%%%%%%%%%%%%%%%%%%%%%%%%%%%%%%%%%%%%%%%%%%%%%%%%%%%%%%%%%
%%%%%%%%%%%%%%%%%%%%%%%%%%%%%%%%%%%%%%%%%%%%%%%%%%%%%%%%%%%%%%%%%g

%%%%%%%%%%%%%%%%%%%%%%%%%%%%%%%%%%%%%%%%%%%%%%%%%%%%%%%%%%%%%%%%%g
\subsection{Correction for Downstream Energy Losses}
\label{sec:downstreamlosses}
\noindent 
In the region analysed in this study, i.e. $\eta=0.687$, 
the electron passes materials with a total thickness
of about $30$ \Xzerox. Therefore
the energy fraction leaking out behind the calorimeter is small.

The amount of energy leaking out of the back of the calorimeter
can be determined from the Monte Carlo simulation.
On average, about $0.35$\% of the initial electron energy with $E= 10$\GeV is deposited
behind the calorimeter,
increasing linearly to $0.45$\% for $E= 245$\GeVx.
Thus, the longitudinal energy leakage
introduces a non-linearity of about $0.1$\% in this energy range.

This effect is corrected on average for each energy point. 

%However,
%it is better to exploit the longitudinal calorimeter segmentation to correct for the leakage
%event-by-event using
%the mean effective shower depth $\av{l}$, defined as the energy weighted barycenter
%over the 
%distance $l$ along the shower axis (measured in $\Xzerox$)
%in the middle of each calorimeter compartment.
%
%The fraction of energy leaking out of the calorimeter as a function of the reconstructed 
%effective shower depth 
%is shown in Fig.~\ref{fig:cleak}. Shown here are Monte Carlo simulations for electron beam energies
%between $10$ and $180$\GeVx. For a given shower effective depth, the leakage fraction is
%almost independent of the initial electron energy. The correction therefore depends only on  $\av{l}$, 
%and not on the electron energy.

%%%%%%%%%%%%%%%%%%%%%%%%%%%%%%%%%%%%%%%%%%%%%%%%%%%%%%%%%
%\begin{figure}[th]
%\begin{center}
%\psfig{figure=figs/cleak.ps,
%bbllx=0,bblly=0,bburx=560,bbury=630,angle=270,clip,width=12.cm}
%\end{center}
%\caption{Energy fraction longitudinally leaking out of the back of the calorimeter as a function of the reconstructed effective
%shower depth. Shown are Monte Carlo simulations (histograms) of electrons with 
%energies $E = 10$\GeVx,  $100$\GeV and  $180$\GeVx. 
%Overlayed as line is a parameterisation.
%\label{fig:cleak}}
%\end{figure}
%%%%%%%%%%%%%%%%%%%%%%%%%%%%%%%%%%%%%%%%%%%%%%%%%%%%%%%%%%%%%%%%%g

\subsection{Correction for the Impact Position within a Cell}
\label{sec:geometry_correction}
\noindent 
Due to the complex structure of the accordion folds, the energy response changes 
as a function of the $\phi$ position of the impinging particle within a cell. 
The main reasons are the varying amount of
passive absorber material %due to the imperfect overlap of the absorbers 
%(see Fig.~\ref{fig:x0vscm}) 
and changes in the electric fields. 
In the $\eta$-direction, a drop of the measured
energy is observed, if the electron does not impinge on the cell centre. 
This effect is due to an incomplete containment of the electron
in the cluster\footnote{
Since the beam for this data sample was mostly covering the central part of the cell, this
effect is small and does not require a correction in this analysis.}.
These effects have been already reported in Ref.~\cite{modul0}.

%To correct for this effect the electron impact position within a cell is reconstructed
%from the calorimeter using:
%\begin{eqnarray}
%\Phi_{calo} =  \frac{\sum_{i=1,N_{cell}} E_i^{vis} \; \Phi_i}{\sum_{i=1,N_{cell}} E_i^{vis} }
%\; \; \; \; 
%\eta_{calo} =  \frac{\sum_{i=1,N_{cell}} E_i^{vis} \; \eta_i}{\sum_{i=1,N_{cell}} E_i^{vis} },
%\label{eq:eta_phi}
%\end{eqnarray}
%where $E_i$ is visible the energy deposited in one cell and $\Phi_i$ and $\eta_i$ are the nominal
%cell positions. To reconstruct $\eta_{calo}$ ($\Phi_{calo}$) only the cells in the first (second) 
%compartment are used.
%The impact position within a cell in cell units
%( $d\eta_{calo}$ and $d\Phi_{calo}$ is then calculated from the known
%cell granularity).

%Because of the finite size of the cells, the energy weighted barycentre is systematically shifted
%towards the centre of the cell. To correct this bias, the $\Phi$ reconstructed with the calorimeter
%is compared to the one measured with the beam chambers  $\Phi_{BC}$.
%The difference $\Phi_{BC} - \Phi_{calo}$ measured as a function of $\Phi_{calo}$ has the form
%of the letter $S$ ($S$-shape), i.e. for a cell impact in the middle of the cell ($\Phi_{calo}=0$) 
%there is no bias, while for cell impact positions below (above) $\Phi_{calo}=0$ there is
%a positive (negative) bias. The form of the distribution can be parameterised with
%$a + b \arctan{((\Phi_{calo} -c)/d)}$, where $a$, $b$, $c$ and $d$ are adjusted parameters.
%This parameterisation is used to correct the calorimeter position measurement.

The electron  $\phi$ impact position within a cell is reconstructed
from the energy weighted barycentre of the second layer.
The bias due to the finite calorimeter cell size is corrected using 
the average difference of the position measurements provided by the calorimeter and by the beam chamber
measured as a function of calorimeter position measurement (``S-shape'').
The corrected $\Phi$ impact position
normalised to middle cell units is called $\Phi_{calo}^{corr}$.
%More details can be found in Ref.~\cite{modul0}. %% only eta !

\begin{figure}[th]
\begin{center}
\psfig{figure=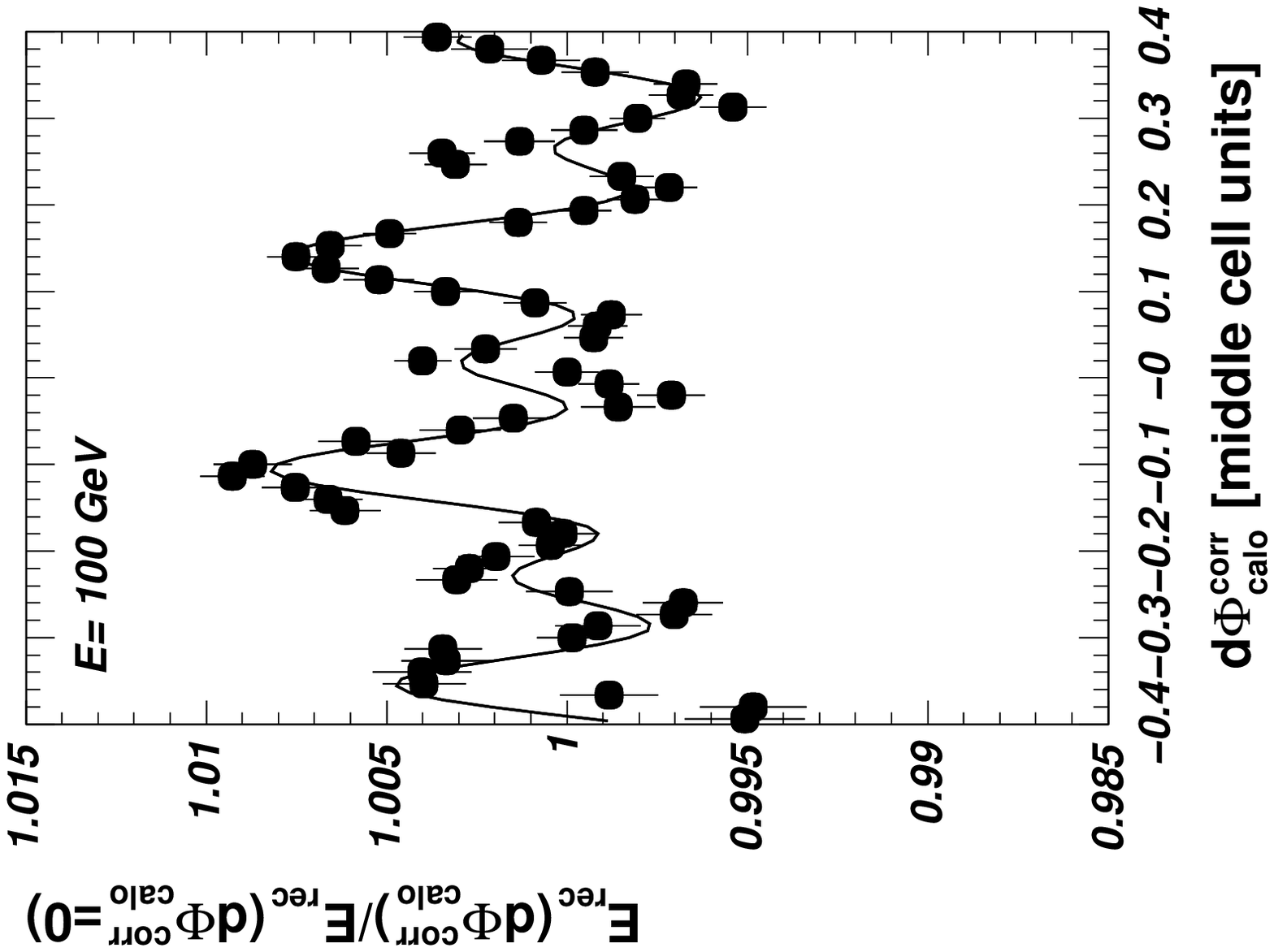,
%bbllx=0,bblly=0,bburx=600,bbury=750,angle=270,clip,
angle=270,width=14.cm}
\end{center}
%\begin{picture}(0,0) 
%\put( 10,10){a)}
%\put( 90,10){b)}
%\end{picture}
\vspace{-0.5cm}
\caption{Correction for the mean energy as a function of the $\phi$ impact
            position within a cell obtained from a data run with electrons at $E=100$\GeVx.
Overlayed as line is a parameterisation.
%         b) Correction for the mean energy as a function of the $\eta$ impact
%            position within a cell
\label{fig:etaphicorr}}
\end{figure}

The meassured dependence of the mean energy on $\Phi_{calo}^{corr}$
is shown in Fig.~\ref{fig:etaphicorr} for an electron beam energy of $E=100 \GeVx$. 
The peak-to-peak modulation for different impact positions is about $1.5\%$. 
%
%The modulation can be parameterised using an energy independent periodical function of the form:
%\begin{eqnarray}
%\label{eq:phi_modulation}
%\frac{E(d\Phi_{calo}^{corr})}{E(d\Phi_{calo}^{corr}=0)} = 
%a + b (\Phi_{calo}^{corr}  - c) + d (\Phi_{calo}^{corr} - c)^2 +
% e \cos{(8 \pi (\Phi_{calo}^{corr} - c))} + f \cos{(16 \pi (\Phi_{calo}^{corr} - c))}
%\end{eqnarray}
%where $a$, $b$, $c$, $d$, $e$ and $f$ are parameters adjusted to the data.
The modulation is parameterised with a function with eight free parameters having 
a  sinusoidal term correcting for the accordion structure and a parabola term
correcting cluster containment effects.
The form of the adjustment is shown in  Fig.~\ref{fig:etaphicorr} as line. 

%The mean energy is altered by this correction
%by less than $0.02$\% for all studied energies. 
%The resolution is not changed for energies up to XXX\GeV and 
%then steadily improves to XXX\% at XXX\GeVx.
The correction is obtained from the data run with an electron beam energy of $100$\GeVx,
where the largest data sample was available, and is applied to all other energies.

%%%%%%%%%%%%%%%%%%%%%%%%%%%%%%%%%%%%%%%%%%%%%%%%%%%%%%%%%
%%%%%%%%%%%%%%%%%%%%%%%%%%%%%%%%%%%%%%%%%%%%%%%%%%%%%%%%%%%%%%%%%g

%A similar effect is observed for the impact position in the $\eta$-direction.
%There the dependence of the mean energy drop towards the edges of the
%cells (see Fig.~\ref{fig:etaphicorr}b). The distribution is asymmetric.
%For cell impact position left from the cell centre a drop of about $1.5\%$ is observed...
%Since most of the events are in the region where the correction is small, no
%improvement of the resolution is obtained. Therefore the correction is not applied.

\subsection{Corrections for Bremsstrahlung Photons lost in the Beam-line}
\label{sec:brems}
\noindent 
Photons produced by Bremsstrahlung of the beam electron in the passive material before the last
trimming magnets, $40$ to $170$~{\rm m} upstream of the detector, 
cannot reach the calorimeter.
Therefore a correction has to be applied to the electron energy measured by the
calorimeter.

The amount of material in this region associated with the NA45 experiment
is not well known and has to be estimated.
In the beam-line, there is about $0.03$ \Xzero of air, 
and about  $0.01$~\Xzero of material from the
beam pipe windows.
In the Monte Carlo simulation this ``far'' material is modelled by a thin spherical shell of 
Aluminium\footnote{The material is simulated as a sphere to ensure the same amount
of material for each $\eta$ direction. In the real experiment the beam-line stays constant and the calorimeter is rotated.
In the Monte Carlo simulation the direction of the beam is changed.}. 
Since the energy lost by Bremsstrahlung leads to a tail on the low energy side
of the reconstructed energy distribution, the amount of material can be estimated
by comparing the tails of the reconstructed energy distribution in Monte Carlo simulations with different
amounts of material and in the data at various beam energies. 
An aluminium thickness of $0.04 \pm 0.01$~\Xzero 
%$3.5$\mm with an uncertainty of $\pm 1$\mm 
has been estimated in this manner. 

The particles produced by Bremsstrahlung in the "far material" are not tracked any further in the
simulation, but their total energy is recorded. 
The correction can then be estimated by looking at the reconstructed %\footnote{
%The energy reconstructed as described in section~\ref{sec:linresults}
%.}
energy with and without the lost energy added to the measured calorimeter energy.
% 
%This effect can influence the low energy tail of the energy distribution and can hence influence
%also the measured mean. 
%For instance, electrons with $180$\GeV ($10$\GeVx)
%that lose $2\%$ of their energy by Bremsstrahlung
%in the ``far material'', arrive with an energy at the calorimeter which is about $3 \sigma$ ($1 \sigma$)
%lower than the nominal beam energy.

If the photon energy is relatively large, the electron energy can be considerably lower than the original
beam electron. Since the beam optics\footnote{In this region there are no bending magnets, but
there are correction dipoles and quadrupoles.} 
is optimised for electrons with the nominal beam
energy, there is a certain probability that the electron will not reach 
the scintillator S3 and S4 defining the beam spot.
This effect has been evaluated using a simulation of the beam-line based on the TURTLE program\cite{turtle}. 
At a beam energy of $10$\GeVx, for an electron having lost $1\%$, $5\%$ or $10\%$ of its energy,
only for
$98\%$, $77\%$ and $50\%$ of the events the electron arrives in the calorimeter.
For a beam energy of $50$\GeVx, 
%for a photon having lost $1\%$, $5\%$ or $10\%$ of its energy only in
the corresponding probabilities are  
$100\%$, $86\%$ and $56\%$. % of the events the electron arrives in the calorimeter.
This correction is applied to the Monte Carlo simulation 
as an event weight for each measured distribution.
%At low electron energies, this effects lowers the total reconstructed energy by about $XXX\%$.

\begin{figure}[th]
\begin{center}
\psfig{figure=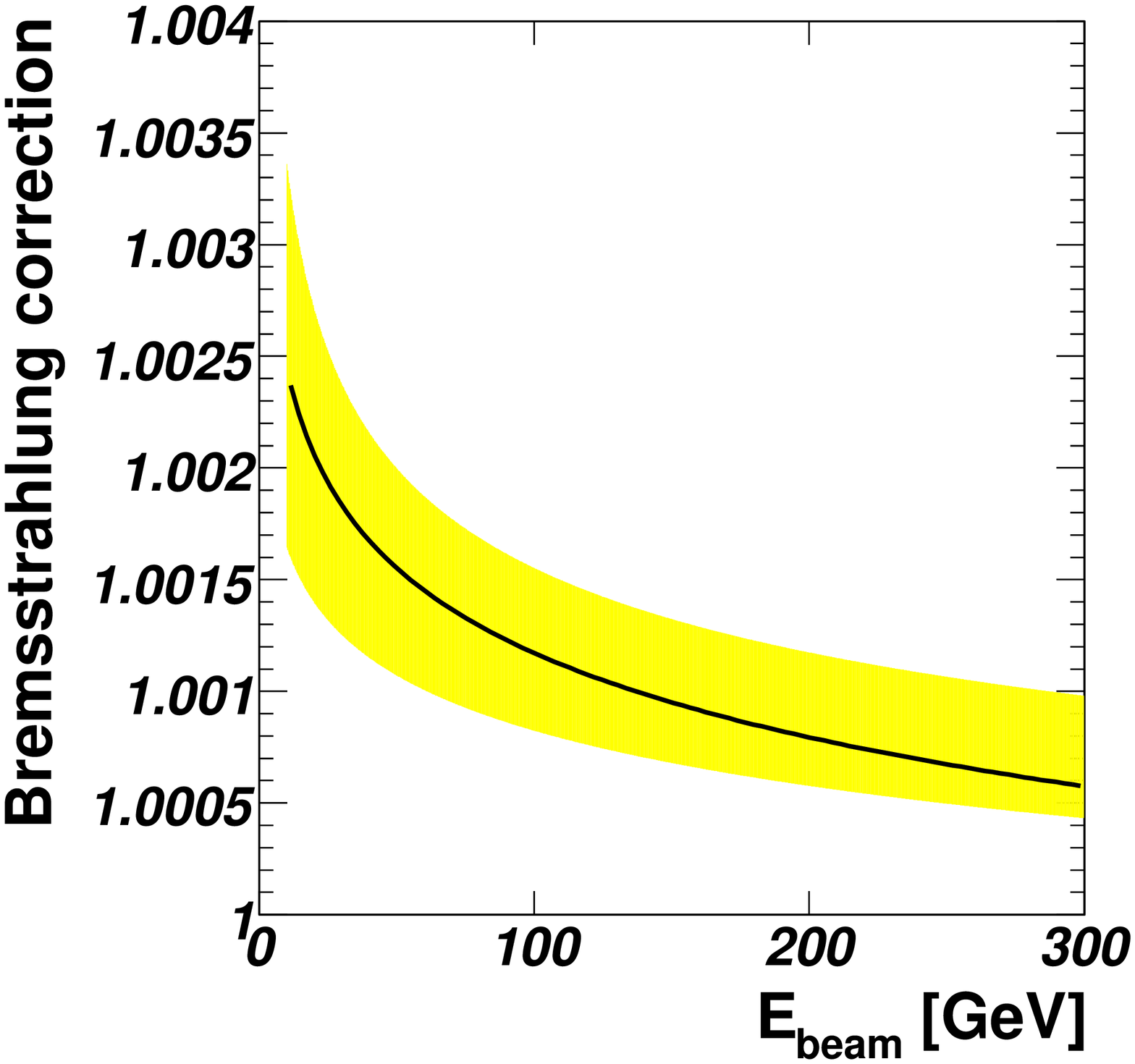,
bbllx=0,bblly=0,bburx=560,bbury=630,angle=270,clip,width=12.cm}
\end{center}
%\vspace{-0.5cm}
\caption{
 Correction for the effect of Bremsstrahlung in the ``far'' material  
 as a function of the electron beam energy.
 The solid line indicates the correction for the standard beam-line set-up. The band
 gives the uncertainty due to variation of the ``far'' material.
\label{fig:cbrems}}
\end{figure}

The correction due to Bremsstrahlung is shown in Fig.~\ref{fig:cbrems}.
For an electron energy of $E = 10$\GeV the peak of the
reconstructed energy distribution is shifted by $0.25$\%,  at  $E = 50$\GeV by $0.15$\%, 
and at  $E = 180$\GeV by $0.09$\%. The non-linearity induced by Bremsstrahlung in the ``far'' material
is therefore about $0.2$\% before correction.

%%%%%%%%%%%%%%%%%%%%%%%%%%%%%%%%%%%%%%%%%%%%%%%%%%%%%%%%%
%%%%%%%%%%%%%%%%%%%%%%%%%%%%%%%%%%%%%%%%%%%%%%%%%%%%%%%%%%%%%%%%%g

\subsection{The Final Electron Calibration Scheme}
\label{sec:final_calib_scheme}
\noindent 
The final electron calibration scheme is a sum of the individual corrections described above:
\begin{eqnarray}
E^{rec} = \left (
a(E) + b(E) \;  E^{vis}_{0} + 
c(E) \; (E^{vis}_{0} \cdot E^{vis}_{1})^{0.5}
+  \frac{1}{d(E) \, f_{samp} } \; \sum_{i = 1,3} E^{vis}_i \right ) \nonumber \\ 
%\cdot f_{\rm brems}(E) 
\cdot f_{\rm cell \, impact}(\Delta \Phi) \cdot (1 + f_{\rm leakage}(E)), 
\;  \;  \;  \;  \;  \;  \;  \;  \;  \;  \;  \;  \;  \;  \;  \;  
\label{eq:Erec}
\end{eqnarray}
where $E^{vis}_{i}$ is the visible cluster energy 
deposited in the $i$th ($i=0,3$) calorimeter compartment,
$f_{samp}=0.18$ is the sampling fraction for an electron with $E=100$\GeV and
the functions $f$ correct (event-by-event) for the effect 
of longitudinal leakage ($f_{\rm leakage}$) (see section~\ref{sec:downstreamlosses})
and of the cell impact position ($f_{\rm cell \, impact}$) 
(see section~\ref{sec:geometry_correction}).
The mean reconstructed energy is in addition corrected for upstream
Bremsstrahlung losses ($f_{\rm brems}$) (see section~\ref{sec:brems}).

%The calibration parameter $a$, $b$, $c$, $d$ are shown in Fig.~\ref{fig:calib_para}.
%The parameters $a$ is expressed in units of \GeVx,
%$b$, $c$ and $d$ have no units.
Since the calibration parameters slightly depend on the energy to be measured,
an iterative procedure is needed to reconstruct the electron energy.

%The calibration parameters are extracted from the Monte Carlo simulation.
%This has the advantage that the various calibrations and corrections can be worked out
%effect by effect, but it requires an excellent
%description of the detector geometry and assumes
%that the description of the physics processes implemented in the \Geant Monte Carlo simulation
%is accurate enough. 
%Uncertainties in the detector description can be estimated and included in the systematic uncertainty.
%The accuracy of the implementation of the physics processes in EM showers is more difficult to evaluate.
%A crude way is to evaluate a systematic uncertainty from the dependence of the cut-off parameter. {\it number to be provided }
%Some information on the validity of the Monte Carlo simulation can be 
%obtained by comparing it to the data.
%This is done in section \ref{sec:data_mc_comparision}.

\section{Comparison of Data and  Monte Carlo Simulations}
\label{sec:data_mc_comparision}
\noindent 
%The shape of the mean reconstructed energy as a function of shower depth $l$ is shown in
%Fig.~\ref{fig:long_prof}. 
%In each calorimeter compartment one measurement is made.
%The data points (closed circles) are well described by the Monte Carlo simulation.
%At the lowest electron beam energy of  $E_{beam} = 10$\GeV
%the shower starts early and most of the energy per radiation
%length is deposited in the first compartment. At the highest electron beam energy
%of  $E_{beam} = 180$\GeV most of the energy per radiation length is deposited in the second compartment.
%In all cases only a small fraction of the energy is deposited in the last calorimeter compartment.
%These measured shower shapes are compatible with an increase of the mean shower depth
%with the logarithm of the electron energy. 
%
%%%%%%%%%%%%%%%%%%%%%%%%%%%%%%%%%%%%%%%%%%%%%%%%%%%%%%%%%
%\begin{figure}[th]
%\begin{center}
%\psfig{figure=figs/long_prof_damc.ps,
%bbllx=0,bblly=0,bburx=560,bbury=630,angle=270,clip,
%width=16.cm}
%\end{center}
%\begin{picture}(0,0) 
%\put(15,  5){c)}  \put(90, 7){d)}
%\put(15, 75){a)} \put(90, 75){b)}
%\end{picture}
%\vspace{-0.5cm}
%\caption{Shape of the reconstructed energy as a function of  
%the shower depth
%for beam energies of $E_{beam}=10$\GeV (a)$, E_{beam}=60$\GeV (b), $E_{beam}=100$\GeV (c) and $E_{beam}=180$\GeV (d).
%Shown are data (closed circles) and a \Geant~Monte Carlo simulation (lines).
%{\it XXXX should one make a ratio plot, i.e. DATA/MC ???
%}
%\label{fig:long_prof}}
%\end{figure}
%%%%%%%%%%%%%%%%%%%%%%%%%%%%%%%%%%%%%%%%%%%%%%%%%%%%%%%%%%%%%%%%%g
%
Since the calibration scheme described above is based on the Monte Carlo simulation, 
it is important to verify that the Monte Carlo simulation reproduces
the total energy distribution,
the energies measured in each layer and the lateral development of the EM shower.

\begin{figure}[th]
\begin{center}
\psfig{figure=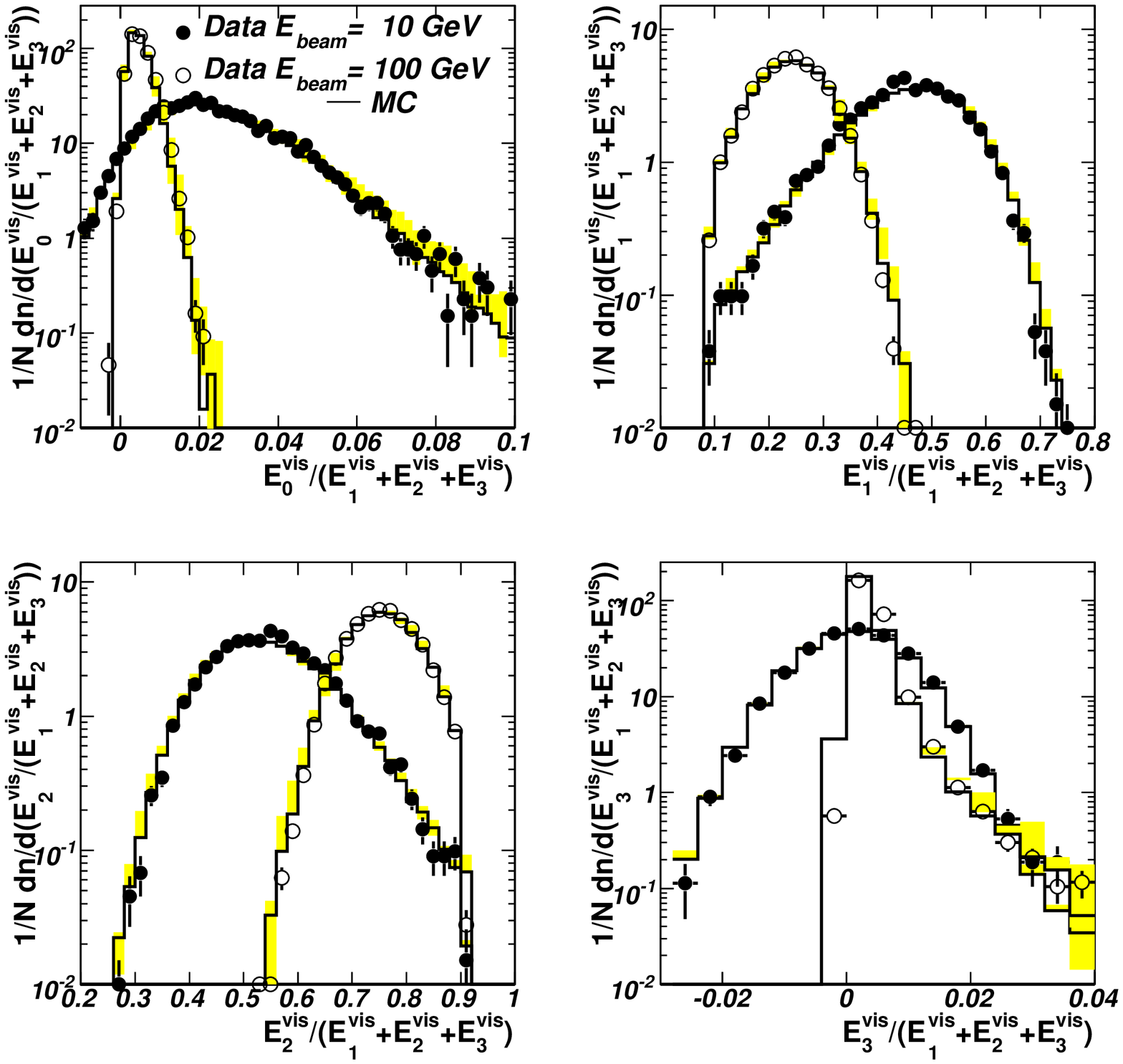,
bbllx=0,bblly=0,bburx=560,bbury=630,angle=270,clip,
width=14.cm}
\end{center}
\begin{picture}(0,0) 
\put(5, 62){a)} \put(75, 62){b)}
\put(5,  1){c)} \put(75,  1){d)}
\end{picture}
%\vspace{-0.5cm}
\caption{
Visible energy fraction distribution for electrons with $E=10$\GeV
and $E=100$\GeV 
in the PS (a) and the first (b), second (c) and third (d)
compartment of the accordion calorimeter. Shown are data (circles)
and a Monte Carlo simulation (line). 
The band indicates the uncertainty
in the Monte Carlo simulation due to the ``far'' material and the material
in front of the PS.
\label{fig:edis100}}
\end{figure}

The mean reconstructed energy in the PS and in the first and second 
compartment of the accordion
is described by the Monte Carlo simulation for all energies within $\pm 2\%$.
In addition, also the shape of the energy distributions within each compartment
are well described. As an example,
the shapes of the visible energy fraction distributions 
are shown 
in Fig.~\ref{fig:edis100} for $E=10$ and $E=100$\GeVx.

%%%%%%%%%%%%%%%%%%%%%%%%%%%%%%%%%%%%%%%%%%%%%%%%%%%%%%%%%
%%%%%%%%%%%%%%%%%%%%%%%%%%%%%%%%%%%%%%%%%%%%%%%%%%%%%%%%%%%%%%%%%g

\begin{figure}[th]
\begin{center}
\psfig{figure=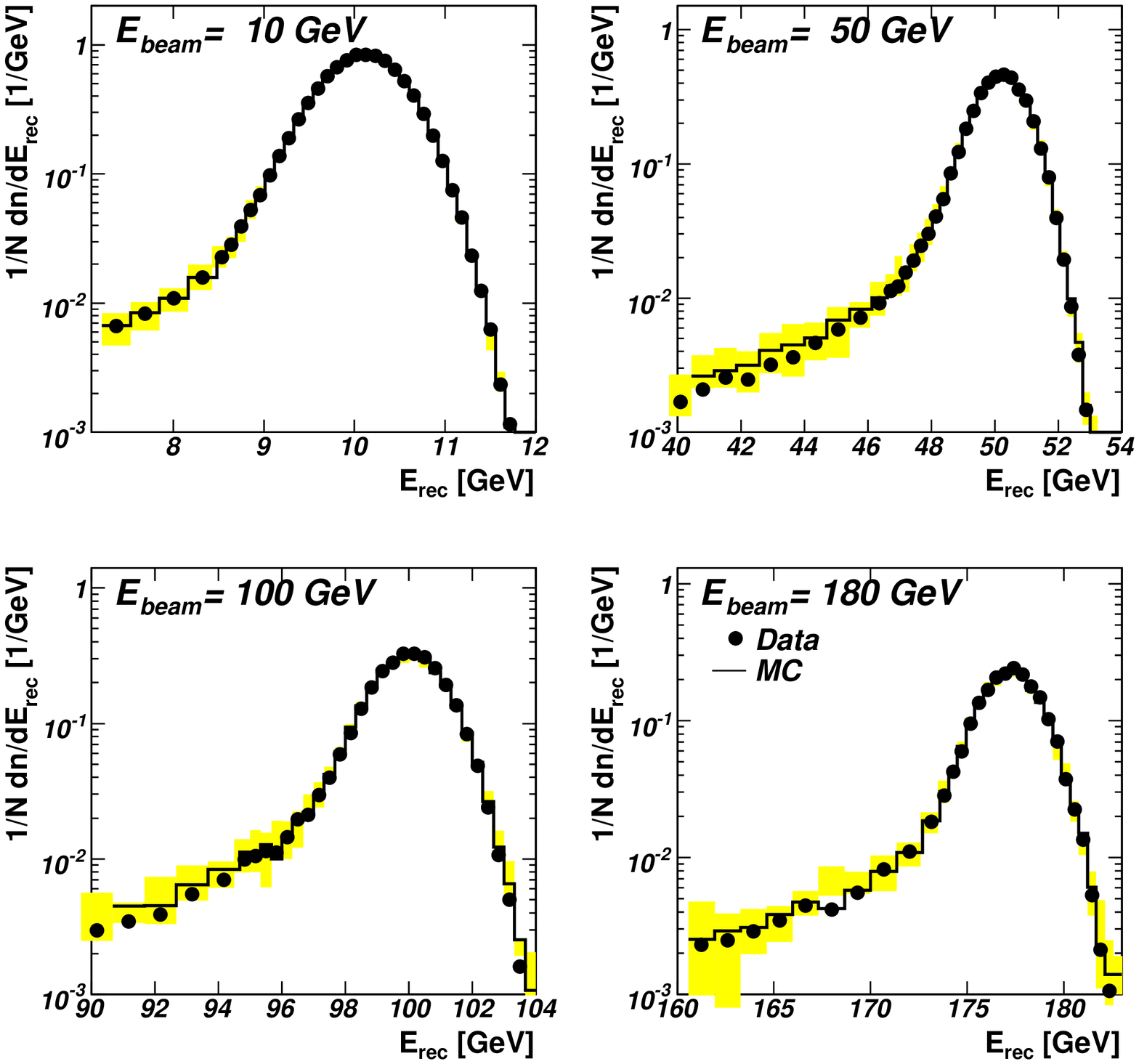,
bbllx=0,bblly=0,bburx=560,bbury=630,angle=270,clip,
width=14.cm}
\end{center}
\begin{picture}(0,0) 
\put(5, 65){a)} \put(75, 65){b)}
\put(5,  0){c)} \put(75,  0){d)}
\end{picture}
%\vspace{-0.5cm}
\caption{
Shape of the reconstructed energy distribution for electrons with 
(a) $E=10$, (b) $50$, (c) $100$ and (d) $180$\GeVx.
Shown are data and a Monte Carlo simulation.
%No corrections for Bremsstrahlung or longitudinal leakage have been applied.
%{\it XXX put here  pions ?}
The band indicates the uncertainty
in the Monte Carlo simulation due to the ``far'' material and the material
in front of the PS.
\label{fig:edis}}
\end{figure}

The distribution of the reconstructed total energy distribution is shown
in Fig.~\ref{fig:edis} for electron beam energies of $10$, $50$, $100$ and $180$\GeVx. 
The beam energy in the Monte Carlo simulation is scaled to the one
in the data. The Monte Carlo simulation gives a good description of the data. 

%At low energy the MC is higher in the low energy tail, 
%at high energy the MC is lower...
%{\it XXXX Why ?? This can be influence by tuning the far material. The problem is that
% the low energy point what less material while the high energy points want more.
% What can can this be ? ..and where it is seen in the single compartments which are
% just summed together ?
%}

%%%%%%%%%%%%%%%%%%%%%%%%%%%%%%%%%%%%%%%%%%%%%%%%%%%%%%%%%
%%%%%%%%%%%%%%%%%%%%%%%%%%%%%%%%%%%%%%%%%%%%%%%%%%%%%%%%%%%%%%%%%

\begin{figure}[th]
\begin{center}
\psfig{figure=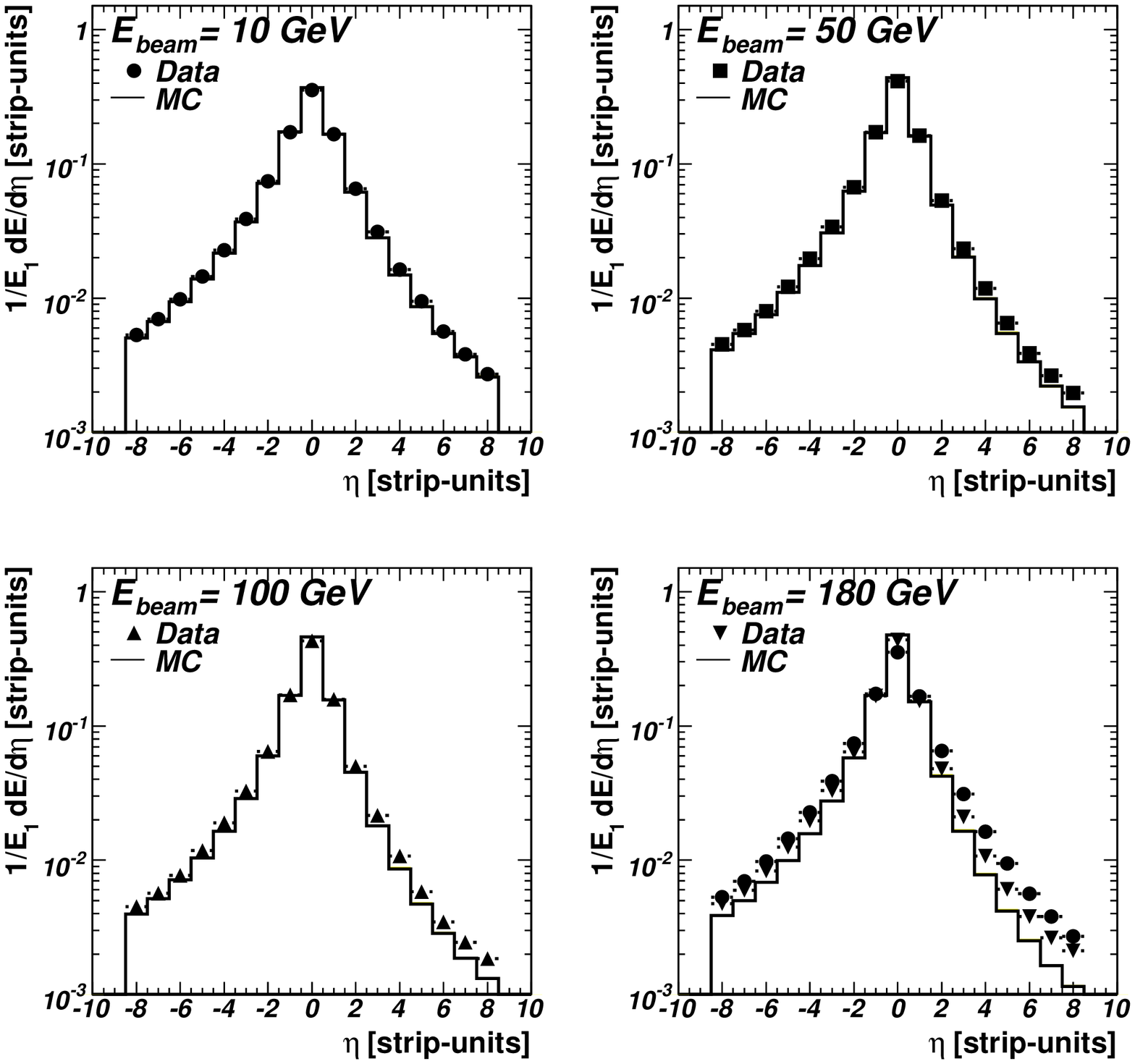,
bbllx=0,bblly=0,bburx=560,bbury=630,angle=270,clip,
width=14.cm}
\end{center}
\begin{picture}(0,0) 
\put(5,  0){c)} \put(75,  0){d)}
\put(5, 65){a)} \put(75, 65){b)}
\end{picture}
%\vspace{-0.5cm}
\caption{Shape of the energy distribution as a function of 
the $\eta$-direction in the
first calorimeter compartment 
for beam energies of (a) $E=10$\GeVx, (b) $60$\GeVx, (c) $100$\GeVx and (d) $180$\GeVx.
Shown are data (closed symbols) and a Monte Carlo simulation (lines).
To illustrate the energy dependence the data with an beam energy of $E=10$\GeV are superimposed in d)
to the data with $E=180$\GeVx.
The systematic uncertainty
in the Monte Carlo simulation due to the ``far'' material and the material
in front of the PS is shown, but is not visible.
\label{fig:lat_prof}}
\end{figure}

The shape of the energy distribution in the $\eta$-direction % radial to the shower axis
measured in the first compartment
is shown in Fig.~\ref{fig:lat_prof} for electron beam energies of
$E=10$, $50$, $100$ and $E=180$\GeVx. The $\eta$-position is calculated
with respect to the shower barycentre in the first compartment and expressed
in units of read-out cells.
%The fine granularity of the first accordion compartment makes such a plot meaningful. 
%Each bin corresponds to one read-out cell ("strip").
%
Due to the compactness of an EM shower the radial extension depends only slightly 
on the beam energy.
The distribution is found to be asymmetric (see also in  Ref.~\cite{strip}). %, 
%i.e. on the left (negative) side more energy is deposited than on the right side. 
%This behaviour has already been observed in Ref.~\cite{strip}.
At low electron beam energies, 
the Monte Carlo simulation gives an excellent description of the data.
In particular, the asymmetry is well reproduced.
At high energies the data distribution is slightly broader than predicted by in the Monte Carlo simulation. 
%This difference might be explained by the cross-talk mentioned in section~\ref{sec:elec_calib}.
%The applied correction only corrects the cell energy, but does not redistribute the energy
%among the cells.

%{\it XXX ...to be investigated: fix a strip number, reject events where maximum is elsewhere,
%  parallel beam in data, but divergent in MC, effect of deta, dphi cuts}

%%%%%%%%%%%%%%%%%%%%%%%%%%%%%%%%%%%%%%%%%%%%%%%%%%%%%%%%%
%%%%%%%%%%%%%%%%%%%%%%%%%%%%%%%%%%%%%%%%%%%%%%%%%%%%%%%%%%%%%%%%%g

%The reconstructed mean energy as a function of the reconstructed energy in the presampler
%and the mean reconstructed shower depth is shown for 
%$E=10$\GeV and $E=180$\GeV in Fig.~\ref{fig:Evsps}.
%The small dependence of the reconstructed energy on these quantities demonstrates
%that the calorimeter system is well calibrated. The response is independent of the
%longitudinal shower fluctuations. The data are described by the Monte Carlo simulation
%within $1\%$.
%
%
%%%%%%%%%%%%%%%%%%%%%%%%%%%%%%%%%%%%%%%%%%%%%%%%%%%%%%%%%
%\begin{figure}[th]
%\begin{center}
%\psfig{figure=figs/Evsps.ps,
%bbllx=0,bblly=0,bburx=560,bbury=630,angle=270,clip,
%width=16.cm}
%\end{center}
%\begin{picture}(0,0) 
%\put(15, 75){a)} \put(90, 75){b)}
%\put(15,  5){c)} \put(90,  5){d)}
%\end{picture}
%\vspace{-0.5cm}
%\caption{Reconstructed energy as a function of the reconstructed
%energy in the presampler (a) and (b) and as a function of the
%shower depth (c) and (d) 
%for electrons with $E=10$\GeV (a,c) and $E=180$\GeV (b,d).
%Shown are data (closed circles) and a Monte Carlo simulation (lines).
%\label{fig:Evsps}}
%\end{figure}
%%%%%%%%%%%%%%%%%%%%%%%%%%%%%%%%%%%%%%%%%%%%%%%%%%%%%%%%%%%%%%%%%g

In conclusion, the Monte Carlo simulation predictions 
are in good agreement with the shower development measured for the data.

\section{Determination of the Pion Contamination using Monte Carlo Simulation}
\label{sec:pion_contamination}
\noindent 
The instrumentation of the H8 beam-line used in the present analysis did not allow 
for a direct measurement of the
pion contamination in the electron beam. Only the signal of the scintillator behind the
calorimeter can be used to reject pions. %A Cerenkov counter was not operational.
The influence of a possible pion contamination  
%on the comparison of data and Monte Carlo simulation 
has therefore to be estimated by comparing the electron and pion energy shapes
predicted by the Monte Carlo simulation to the one measured in the data.

%For one electron run with a beam energy of $50$\GeVx, no on-line
%pion veto based on a scintillator behind the calorimeter was applied.
%In this run, all pions in the electron beam can be measured and the ability
%of the Monte Carlo simulation to describe the interaction of pions in the
%EM calorimeter can be tested. From the shape of the measured energy distribution 
%the fraction of pions under the electron peak can be estimated.
%
%The reconstructed energy spectrum is shown in Fig.~\ref{fig:cpion2}.
%An appropriate mixture of the electron and the pion simulation is able to describe it.
%The dotted line illustrates a pion simulation, the dashed one an electron simulation.
%The peak at very low energies is due to pions traversing the calorimeter
%as minimally ionising particles, the wide distribution around $25$\GeV is due
%to pions undergoing a strong interaction and leaving only part of
%their energy in the EM calorimeter. The tail towards high energies
%is caused by pions depositing most of their energy fully electromagnetically.

Since the \LAr calorimeter has a thickness of about one interaction length,
most pions deposit only a fraction of their incident energy.
%Therefore most of the pions with the same beam energy as the electrons under study,
%do not influence the measurement of the electron energy. 
However, a small fraction
of them can deposit most of their energy in the \LAr calorimeter. For instance,
at $E=10$\GeV about $3$\% of the pions deposit more than  $8$\GeV
in the \LAr calorimeter. 
Since these pions can influence the measurements of the electron energy,
their fraction has to be determined.
%Since the shape of their energy distribution is decreasing towards
%the beam energy, the low energy tail can be used to determine
%the pion fraction in the data.
%
%Since the Monte Carlo simulation gives a reasonable description of the
%reconstructed energy spectrum, the low energy tail of the measured
%energy distribution can be used to study the effect of pion contamination
%also for the other energies where only the pions which are absorbed in
%the EM calorimeter have been recorded. 

\begin{figure}[th]
\begin{center}
\psfig{figure=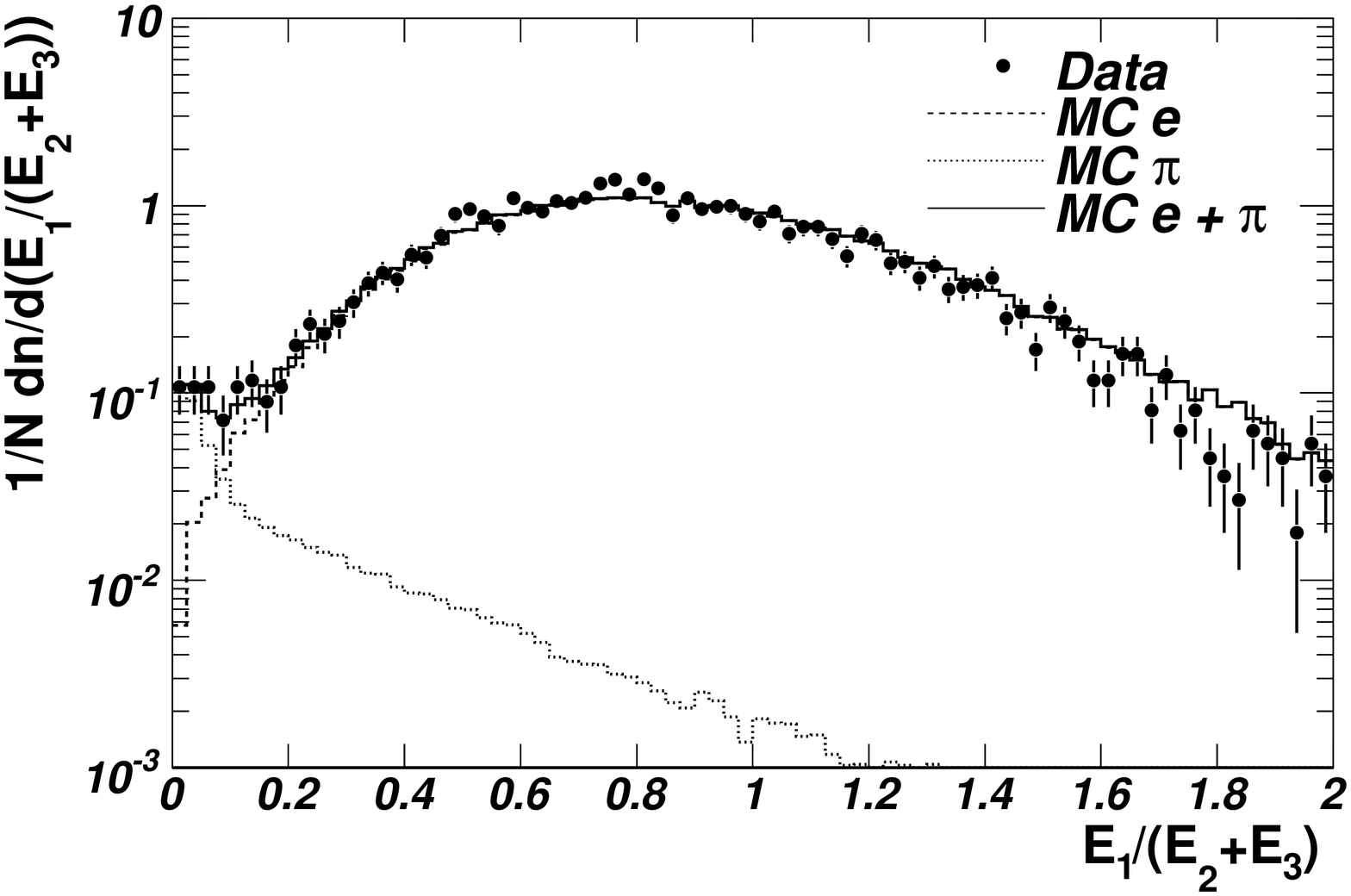,
bbllx=0,bblly=0,bburx=570,bbury=770,angle=270,clip,width=12.cm}
\psfig{figure=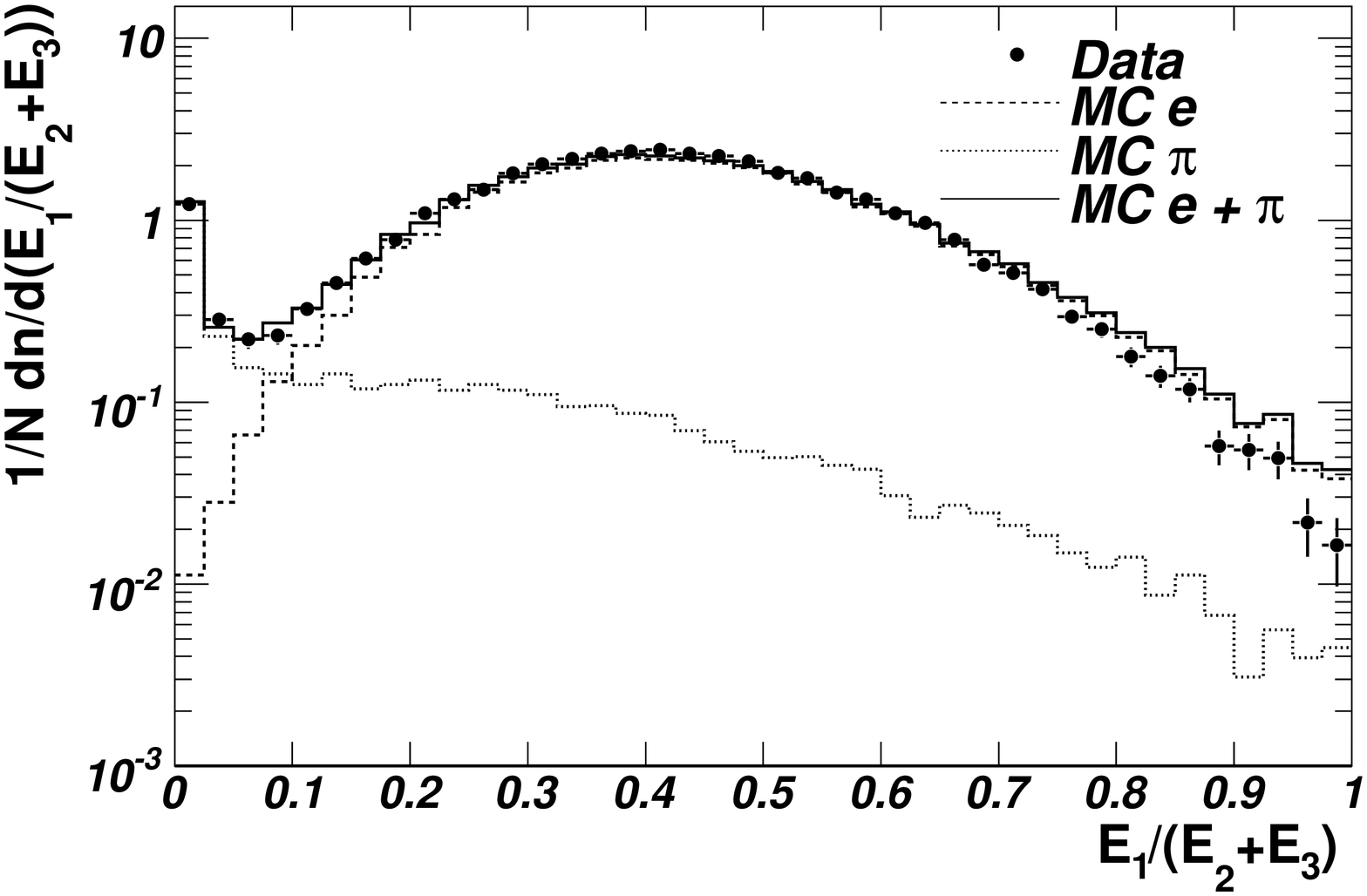,
bbllx=0,bblly=0,bburx=570,bbury=770,angle=270,clip,width=12.cm}
\end{center}
\begin{picture}(0,0) 
\put(10,  90){a)} \put(10,  5){b)}
\end{picture}
\vspace{-0.5cm}
\caption{
Shape of the ratio of the first ($E_1$) over the sum of the second ($E_2$) 
and of the third ($E_3$) calorimeter compartment.
Shown are data (closed circles) 
with  $E = 10$\GeV (a) and $E = 50$\GeV (b)
and Monte Carlo simulations 
for electrons (dashed), pions (dotted) and 
an appropriate mixture of electrons and pions (solid). 
\label{fig:cpion}}
\end{figure}

%The fraction of pions in the electron beam can be estimated using Monte Carlo simulations. 
Pions which deposit a lot of energy in the \LAr calorimeter
interact on average later in the
\LAr calorimeter than do electrons. %, i.e. they deposit less energy in the first calorimeter
%compartment ($E_1^{rec}$) and more 
%in the second ($E_2^{rec}$) and the third ($E_3^{rec}$) ones than electrons.
Compared to electrons of the same beam energy
they deposit therefore less energy in the first compartment and more in the second and third ones.
The ratio $E_1^{vis}/(E_2^{vis}+E_3^{vis})$ is 
shown in Fig.~\ref{fig:cpion}a for $E = 10 \GeV$ and Fig.~\ref{fig:cpion}b for $E = 50 \GeV$
for data and an appropriate mixture of electrons and pions.
The Monte Carlo simulation is able to describe the data. 
%The  pion fraction can therefore be determined.
The fraction of pions in the electron beam 
is determined for each energy
and varies from $2$\% at $10$\GeV to $22$\% at $180$\GeVx.
The effect on the electron energy measurement will be discussed in section~\ref{sec:syslinresults}.

%%%%%%%%%%%%%%%%%%%%%%%%%%%%%%%%%%%%%%%%%%%%%%%%%%%%%%%%%
%\begin{figure}[th]
%\begin{center}
%\psfig{figure=figs/cpion2.ps,
%bbllx=0,bblly=0,bburx=560,bbury=700,angle=270,clip,width=12.cm}
%\end{center}
%%\vspace{-0.5cm}
%\caption{Shape of the reconstructed energy for data and Monte Carlo simulations 
%        for electrons at  $E=50$\GeVx.
%Shown are electron data (closed circles) 
%and \Geant~Monte Carlo simulations 
%for electrons (dashed), pions (dotted) and a mixture of electrons and pions (solid). 
%\label{fig:cpion2}}
%\end{figure}
%%%%%%%%%%%%%%%%%%%%%%%%%%%%%%%%%%%%%%%%%%%%%%%%%%%%%%%%%%%%%%%%%g

%%%%%%%%%%%%%%%%%%%%%%%%%%%%%%%%%%%%%%%%%%%%%%%%%%%%%%%%%
%%%%%%%%%%%%%%%%%%%%%%%%%%%%%%%%%%%%%%%%%%%%%%%%%%%%%%%%%%%%%%%%%g

%\newpage 
\section{Linearity and Resolution Results}
\label{sec:results}
\subsection{Linearity Results}
\label{sec:linresults}
\noindent 
%The mean energy is obtained by adjusting a Gaussian to the reconstructed energy
%distribution. This is done in two steps:
%First, a Gaussian is adjusted in a relatively wide range around the mean reconstructed
%energy. The standard deviation from this fit is used to determine the range for the final
%adjustement of a Gaussian to the reconstructed energy distribution.
%Two standard deviations for the low energy side and three standard deviations
%for the high energy side are used. This is the maximal possible fit range
%where the $\chi^2$ per degree of freedom is one. To determine the uncertainty
%due to the choosen fit range, the range is restricted to $1.5$ and extended to $2$ standard
%deviations.
%
The mean energy is obtained by fitting a Gaussian to the reconstructed energy
distribution within
two standard deviations for the low energy side and three standard deviations
for the high energy side\footnote{This is the maximal possible fit range
where the $\chi^2$ per degree of freedom is one.}. To determine the uncertainty
due to the chosen fit range, 
results are also considered where 
the range of the low energy side
is restricted to $1.5$ and extended to $2.5$ standard deviations.

\begin{figure}[th]
\begin{center}
\psfig{figure=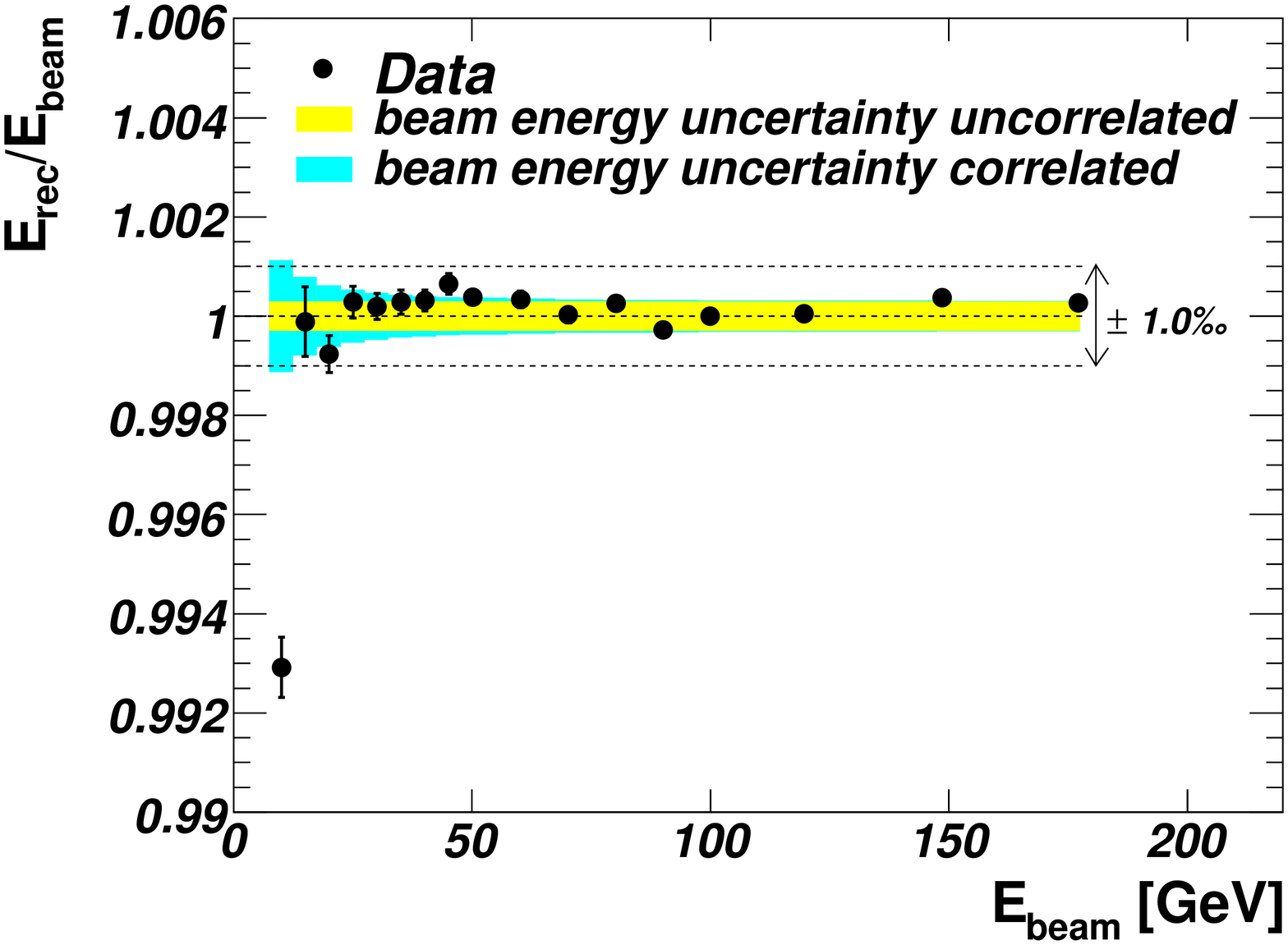,
bbllx=30,bblly=40,bburx=700,bbury=730,angle=270,clip,
width=14.cm}
\end{center}
\vspace{-2.5cm}
\caption{
Ratio of the reconstructed electron energy to the beam energy 
as a function of the beam energy. All points are  normalised to the value measured at $E = 100$\GeVx.
The inner band illustrates the uncorrelated uncertainty of the beam energy measurement; 
in the outer band the correlated
uncertainty is added in quadrature to the inner band.
\label{fig:elin}}
\end{figure}

The mean reconstructed energy divided by the beam energy is shown in Fig.~\ref{fig:elin}.
The error bars indicate the statistical uncertainty as obtained by the fit procedure.
Since the absolute calibration of the beam energy is not precisely known, all points
are normalised to the value measured at $E = 100$\GeVx.
The inner band represents the uncorrelated uncertainty on the knowledge of the beam energy,
while the outer band shows in addition the correlated uncertainty added in quadrature 
(see section~\ref{sec:beam}).
For energies $E > 10$\GeVx, all measured points are within $\pm 0.1$\%. 
The point $E = 10$\GeV is lower by $0.7$\% with respect to the other measurements.

%%%%%%%%%%%%%%%%%%%%%%%%%%%%%%%%%%%%%%%%%%%%%%%%%%%%%%%%%%%%%%%%%
%%%%%%%%%%%%%%%%%%%%%%%%%%%%%%%%%%%%%%%%%%%%%%%%%%%%%%%%%%%%%%%%%g

\subsection{Systematic Uncertainties on the Linearity Results}
\label{sec:syslinresults}
\begin{figure}[th]
\begin{center}
\psfig{figure=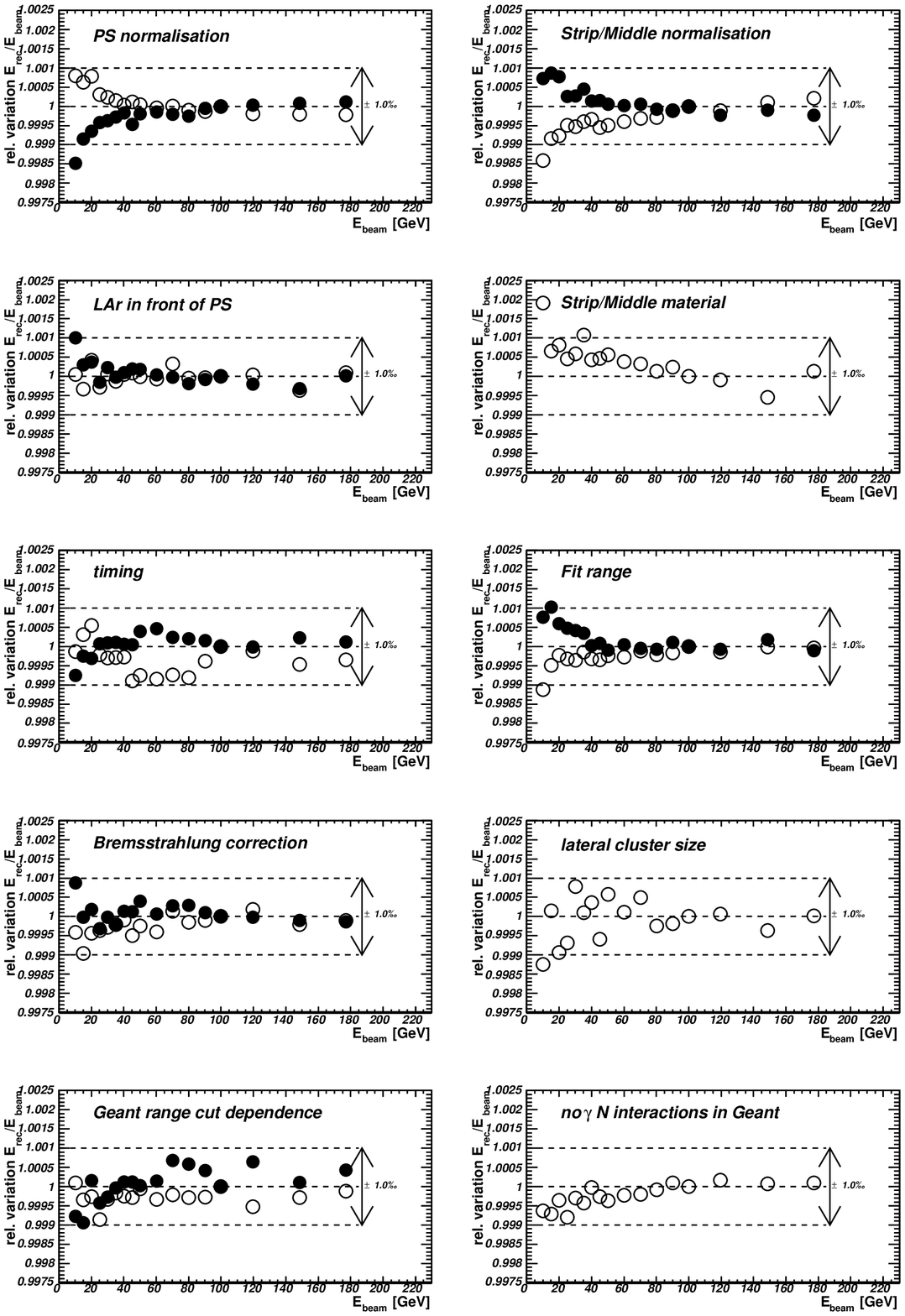,
%bbllx=0,bblly=0,bburx=560,bbury=630,angle=270,clip,
width=14.cm}
\end{center}
\begin{picture}(0,0) 
%\put( 5,175){a)} 
%\put(75,175){b)}
\put( 5,168){a)}  
\put(75,168){b)}
\put( 5,128){c)}  
\put(75,128){d)}
\put( 5, 88){e)}  
\put(75, 88){f)}  
\put( 5, 48){g)}  
\put(75, 48){h)}
\put( 5,  5){i)}  
\put(75,  5){j)}  
\end{picture}
\vspace{-0.5cm}
\caption{Sensitivity to systematic effects on the electron energy measurements:
 a) the normalisation of the PS, 
 b) the relative normalisation of the strips and the middle, 
 c) the amount of \LAr in front of the PS,
 d) the amount of material between the PS and the strips
 e) the event arrival time with respect to the data acquisition clock,
 f) the range where the Gaussian fit is performed,
 g) the Bremsstrahlung correction (amount of "far" material),
 h) the lateral shower description,
 i) the dependence on the range cut used in the simulation
% j) the deformation of the calorimeter cells due to the gravitational force
 j) the presence of photon nucleon reactions in the simulation,
% l) the $\Phi$-impact point correction within a cell.
Shown is the relative variation of the reconstructed to the beam electron energy.
All points are normalised to the value measured at  $E=100\GeVx$.
The closed and open circles show the variations of the corresponding systematic uncertainties
(see text) relative to the default.
\label{fig:syst}}
\end{figure}

\noindent 
The systematic uncertainties induced by various effects on the reconstructed electron
energy  are shown in Fig.~\ref{fig:syst}. In order to evaluate the size of some of
the systematic uncertainties, dedicated Monte Carlo simulations have been produced
to calculate new sets of calibration parameters. These samples were typically smaller
than the default one.

The uncertainty on the current to energy conversion factor (see section~\ref{sec:fieps})
of the PS has been studied using the $\chi^2$-distribution of the visible
energy distribution for data and Monte Carlo simulations for all energy points.
The uncertainty is estimated by the scatter for different energies.
The same procedure has been repeated by studying the dependence of the mean reconstructed energy
on the PS energy in the data and in the Monte Carlo simulations.
A consistent result has been found.
Since the relative contribution for the PS is larger at low energies, the systematic
uncertainty rises towards low energies (see  Fig.~\ref{fig:syst}a). 
While  the systematic uncertainty is negligible at  $E = 180$\GeVx, 
it reaches  about $0.1$\% at  $E = 10$\GeVx. 

The uncertainty due to the relative normalisation difference between the first and the
second compartments (see section~\ref{sec:fieps}) is shown in Fig.~\ref{fig:syst}b.
This effect biases the energy measurement by up to about $0.1$\%, mostly at low energies.

The systematic uncertainty arising from the incomplete knowledge of the amount of \LAr
between the PS and the \LAr excluder in front of it (see section~\ref{sec:g4})
is shown in Fig.~\ref{fig:syst}c. It introduces an uncertainty of about $0.05$\%.
Again, low energies are most affected.

Fig.~\ref{fig:syst}d shows the effect of adding ad hoc $0.02$~\Xzero additional material
between the PS and the first compartment. The relative variation of the
reconstructed beam electron energy is slowly decreasing from low to high energies 
The effect amounts to about $0.1$\% at low energies. At $E = 10$\GeV even $0.2\%$ is found.

%Contrary to the case of the data taking of the finalised ATLAS experiment 
%where the protons are collided every $25$\nsx, in the test-beam
%the electrons arrive at any time with respect to the $25$\ns clock used by the
%data acquisition system. In this respect the energy reconstruction is more complicated
%in the test-beam than it will be in ATLAS, since it depends on the event timing. 
%For instance, while in ATLAS only one set of optimal filtering coefficients
%will be needed, in the test-beam $25$ different sets for each\ns are used
%to reconstruct the energy. However, some of the time dependence of the energy
%reconstruction is already introduced during the data taking. 
%The decision where the electronics switches from high to medium gain is based
%on the signal measured in a fixed time sample. The electronics is adjusted such that
%for most events the signal amplitude lies in the middle of this time
%sample. However, when the electrons arrive a bit later or earlier the amplitude
%is shifted left or right. This on-line procedure introduces therefore
%a dependence of the reconstructed energy on the electron arrival time.

As explained in section~\ref{sec:calib}, according to the amplitude of a predefined sample 
in the medium gain, a selection of the gain to be digitised is done. In the test-beam 
electrons arrive at any time with respect to the $25$\ns clock used by the
data acquisition system. When this sample is not at the peak of the signal, 
it happens around the threshold that the cell is digitised in the high gain while it
should have been done in the medium gain. The fraction of such events depends on the trigger phase with
respect to the 40 MHz clock. Moreover, in the calibration procedure
no continuity of the reconstructed amplitude near the overlap region of high
and medium gain
has been imposed. A combination of these two effects can induce a change of the 
reconstructed energy especially in the electron range from $40$ to $80$\GeVx. This effect was studied by selecting 
events sampled near the peak ($10 < t_{tdc} < 20 $\nsx) and outside this time window 
($t_{tdc} < 10 $\ns or $t_{tdc} > 20 $\nsx).
In ATLAS a timing adjustment will be performed such that the maximum sample is near the maximum
of the signal, which corresponds to $15$\ns in the testbeam data taken
in asynchronous mode.
As shown in Fig.~\ref{fig:syst}e, the largest effect is about  $\pm 0.1 \%$ at $60$\GeVx.  
%In Fig.~\ref{fig:elintdc} the linearity performance for events sampled near the peak is presented.

%In Fig.~\ref{fig:syst}d the uncertainty introduced by this effect is shown.
%Shown %as closed  circles%
%are events where the
%electron arrives within $10 < t_{tdc} < 20 $\nsx, i.e. in the middle of the
%$25$\ns window %. The open circles show the uncertainty introduced by events arriving 
%and events arriving
%at the edges, i.e. outside $10 < t_{tdc} < 20$\nsx.
%In the region around $50$\GeV where the highest energetic cell switches gain
%the uncertainty is largest. The maximal uncertainty is $0.1$\%.

The uncertainty introduced by restricting or extending the fit range 
of the Gaussian to the reconstructed energy distribution (see section~\ref{sec:linresults})
is shown in Fig.~\ref{fig:syst}f. At low energies the uncertainty reaches $0.1$\%,
above $E > 60$\GeV it is negligible.

The bias introduced by the uncertainty of the "far" material 
and correspondingly of the Bremsstrahlung correction
(see section~\ref{sec:brems}) is shown in Fig.~\ref{fig:syst}g.
The resulting uncertainty is about $0.05\%$ (up to $0.1\%$ at two energies).

To test the influence of an incomplete description of the low energy
tail laterally to the shower axis, the whole analysis is repeated using
a $5{\rm x 5}$ instead of a $3{\rm x 3}$ cluster. 
The change with respect to the standard analysis
is shown in Fig.~\ref{fig:syst}h). The uncertainty is about $0.1\%$.

Fig.~\ref{fig:syst}i) shows the effect of using different range cuts in the
Monte Carlo simulation. The default value of $20${\rm $\mu$m} is decreased
to $10${\rm $\mu$m} and increased to $100${\rm $\mu$m}.
Although visible energy and sampling fraction significantly change in the Monte Carlo simulation, 
the linearity remains constant within about $0.05\%$ ($0.1\%$ in exceptional cases).

In the default Monte Carlo the small deformation of the calorimeter cells
in the gravitational field of the earth
is modeled ("sagging"). 
%The deviation to the case where this deformation is neglected 
%is shown in Fig.~\ref{fig:syst}j). 
This effect introduces at most a change of $0.05\%$.

In an electromagnetic shower hadronic interactions of mainly photons with
nucleons can lead to deposited energy that can not be measured in the calorimeter
(nuclear excitation etc.) or can produce particles that escape detection 
(neutron, neutrinos etc.). According to the Monte Carlo simulation 
on average about $0.4$\% of the energy can not be measured in the calorimeter. 
However, the energy dependence of this effect is small.
While at high energies the relative variation of a Monte Carlo simulation
with photon nucleon interactions switched off to the default case
is constant, at low energies it is about  $0.995-0.999$.
This is shown in Fig.~\ref{fig:syst}j).

The correction for the modulation of the reconstructed energy on the
$\phi$-impact position within a cell (needed to improve the energy resolution)
does not change the mean reconstructed energy within $0.05\%$. % (see Fig.~\ref{fig:syst}l).  
%For smaller energies the correction introduces a small change of the mean reconstructed energy, for example
%$0.1\%$ at $E=15\GeV$ 
%This means that this correction has a small energy
%dependence\footnote{Because of the limited statistics of the data sample
%this correction has been derived at $E=100\GeVx$, where the largest data set
%was available, and is applied to all energies.}.

%%%%%%%%%%%%%%%%%%%%%%%%%%%%%%%%%%%%%%%%%%%%%%%%%%%%%%%%%%%%%%%%%
%%%%%%%%%%%%%%%%%%%%%%%%%%%%%%%%%%%%%%%%%%%%%%%%%%%%%%%%%%%%%%%%%

%The more energy they distribute, the more they deposit
%their energy electromagnetically.
%The electron energy measurements might be systematically reduced by a possible
%pion contamination, since pions lead to an energy distribution 
%in the \LAr calorimeter which decreases towards the beam energy. 
To estimate how the reconstructed electron energy is biased
by a possible pion contamination, 
the fraction of pions with a large energy deposit in the \LAr
is determined from the Monte Carlo simulation
(see section~\ref{sec:pion_contamination}) and the shift of the reconstructed
mean electron energy in the Monte Carlo simulation is calculated.
%
%As an example the shape of the measured energy distribution
%is shown in Fig.~\ref{fig:cpion2}. 
%In the region $40 < E_{rec} < 50$\GeV
%$0.4\%$ of the events are pions. However, no shift of the mean reconstructed
%electron energy is introduced. This would be only the case, if the
%pion contamination was ten times larger.
The shift is negligible at all energies.
%
%This is the case for pions with $E_{rec} > 7$\GeV (dashed line) as well as  
%for pions with $E_{rec} > 9.5$\GeV (dotted line).
%Electrons at $E_{beam}=10$\GeV  deposit about
%the same amount energy in the first than they do in the second compartment.
%The Monte Carlo simulation for electrons describes the data well. In particular,
%the region where most of the pion contamination is expected is well described.

%To further test  the influence of a possible pion contamination in the data,
In addition, 
data events with $E_1^{vis}/(E_2^{vis}+E_3^{vis}) > 0.1$ can be selected. Such a cut does not
shift the reconstructed energy in the Monte Carlo simulation of electrons
and does not change much the expected pion energy distribution.
%This data sample is expected to have about half of the pions removed.
The reconstructed energy in the restricted data sample, where about half of the pions
are expected to be removed, does not change. % with respect to the full data sample.
% one can conclude, that the effect of pion
%contamination is negligible for the energy measurement.

To ensure a precise measurement of the electron energy, the energy dependence
of the calibration parameters has to be taken into account.
Since the initial electron energy is not known a-priori, an iterative
procedure has to be applied. Starting from the measurement in the accordion calorimeter
and using an average calibration parameter for the accordion calorimeter 
(parameter $d$ in eq.~\ref{eq:Erec}), the energy is reconstructed.
With this first energy estimate, the calibration corrections are evaluated.
This procedure is evaluated until the reconstructed energy does not change
significantly. Already after two iterations an accuracy of better than $10^{-5}$ is achieved.

\subsection{Interpretation of the Linearity Results}
\label{sec:interlinresults}
\noindent 
To quantify the non-linearity of the measured data points for $E > 10 \GeVx$,
a first order polynomial is fitted. The resulting slope ($a_1$) is compatible with zero.
%%The result is shown in Fig.~\ref{fig:elintdc} (solid line)
%and is within the systematic uncertainty on the beam energy.
%The slope $a_1 = (-4  \pm 9) \cdot 10^{-6} \; \GeVinvx$, is compatible with zero.
%implying that  lower energies are overestimated with respect to the higher energies.
%
%
%
%
%When instead of a normal polynom a first order Legendre-polynoms are fitted,
%the slope is $l_1 = -1.2  \pm 0.8 \cdot 10^{-4} \; \GeVinvx$. Since 
%he Legendre-polynoms are orthogonal, the slope stays approximately unchanged
%when higher orders are fitted in addition. 
%
When the analysis is repeated for each systematic uncertainty, slopes 
in the range $a_1 = \pm 5 \cdot 10^{-6}  \GeVinv$ are obtained.
All systematic uncertainties combined in quadrature give an uncertainty
of  $\pm 9 \cdot 10^{-6}  \GeVinv$.
%The systematic
%uncertainty on the fit with the Legendre Polynoms is 
%$l_1 = - 4 \cdot 10^{-4} - 0.2 \cdot 10^{-4} \; \GeVinvx$. Therefore within the systematic
%uncertainties no non-linearity is observed.
%In conclusion, the overall non-linearity is smaller than $5 \cdot 10^{-4} \; \GeVinvx$.
%
Based on purely statistical uncertainties, the $\chi^2$ per degree of freedoms
for a linear fits to the data points
is $\chi^2/ndf = 2.7$. This, together with the fact that the pull
distribution is not Gaussian (RMS is about 1.5), indicates that the measured data points 
are not fully compatible with a straight line and that systematic uncertainties
affect the linearity.

%When the measurement is restricted to events arriving
%in the middle of the time interval with respect to the data acquisition clock,
%the slope is also compatible with zero
%(see open circles and dashed line in Fig.~\ref{fig:elintdc}), but
%in this case the  $\chi^2/ndf = 1$ and the pull distribution 
%is compatible with a Gaussian with a mean at zero and 
%a standard deviation of $1.3 \sigma$.
%
%%%%%%%%%%%%%%%%%%%%%%%%%%%%%%%%%%%%%%%%%%%%%%%%%%%%%%%%%%%%%%%%%
%\begin{figure}[th]
%\begin{center}
%\psfig{figure=figs/celintdc.ps,bbllx=0,bblly=0,bburx=580,bbury=830,angle=270,clip,width=14.5cm}
%\end{center}
%\vspace{-0.5cm}
%\caption{
%Ratio of the reconstructed electron energy to the beam energy 
%as a function of the beam energy. The closed circles indicate the results
%of using the full data set, the open circles are obtained by restricting
%the measurements to events with $10< t_{tdc} <20$ \nsx, for which 
%the electron arrives close to the  data acquisition clock.
%All points are  normalised to the value measured at $E = 100$\GeVx.
%A fit of a linear function to the data points displayed as 
%closed (open) circle is shown as solid (dashed) line.
%The inner band illustrates the uncorrelated uncertainty of the beam energy measurement;
%in the outer band the correlated
%uncertainty is added in quadrature to the inner band.
%\label{fig:elintdc}}
%\end{figure}
%%%%%%%%%%%%%%%%%%%%%%%%%%%%%%%%%%%%%%%%%%%%%%%%%%%%%%%%%%%%%%%%%g

\begin{figure}[th]
%\vspace{0.5cm}
\begin{center}
\psfig{figure=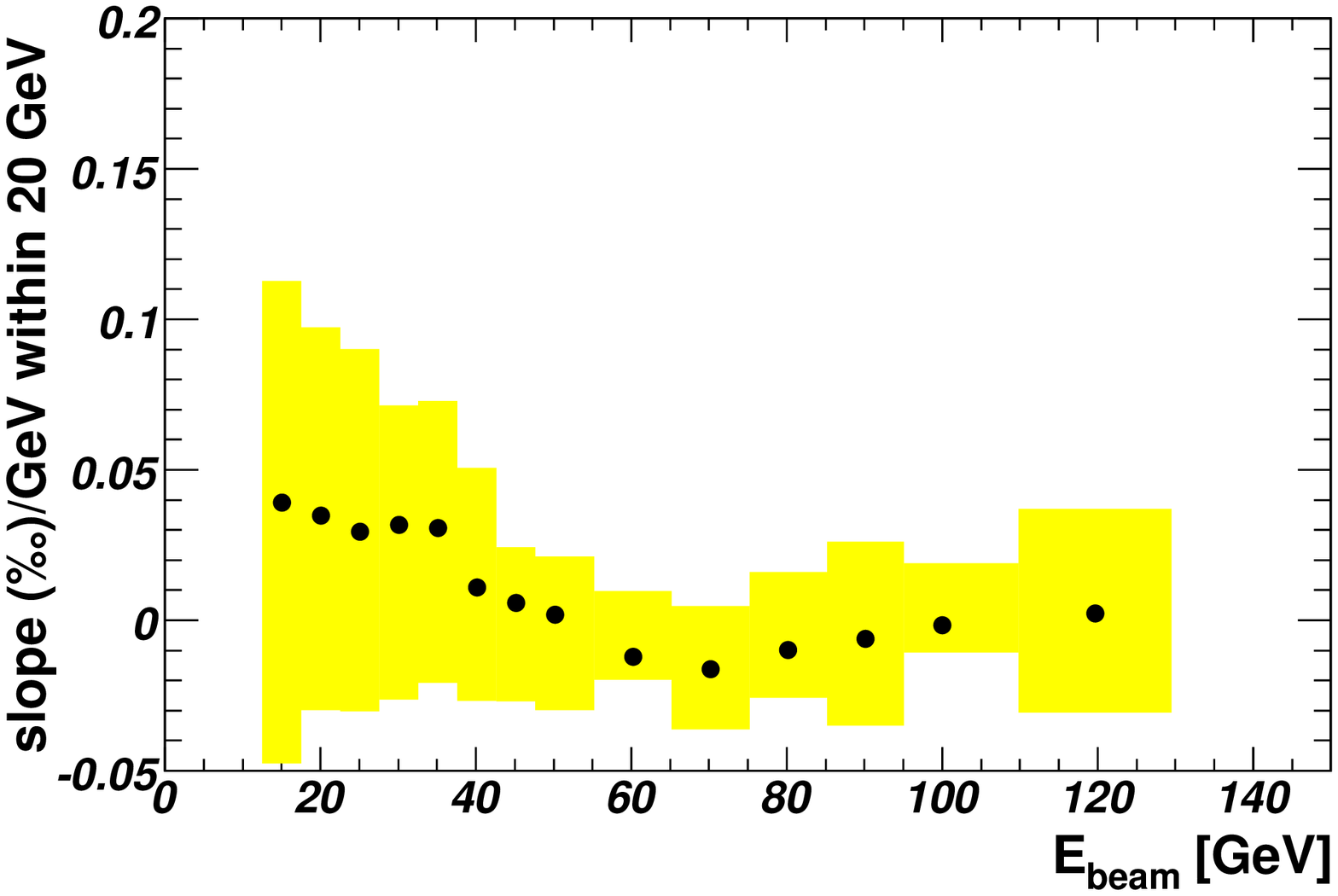,width=14.cm}
\end{center}
%\vspace{-0.3cm}
\caption
{Slope calculated for the energy measurements within $20$\GeV around a given 
 electron energy. The point represent the default result, while
 the band illustrates the slopes obtained from the systematic variations.
\label{fig:localslopes}}
\end{figure}

In practical applications like the measurement of the $W^\pm$-boson mass, 
the shift of the measured (transverse) energy spectrum with respect to a
reference reaction like that from the $Z^0$-boson needs to be understood.
Since the transverse energy distribution is roughly peaked at half of the boson mass and
slowly decreases towards lower transverse energies, one is 
interested in the control of the linearity within a few\GeVx.
To estimate the size of local non-linearities 
for each energy measurement the local slope is calculated from the measurement
which have a beam energy difference smaller than $20$\GeVx.
%So, for the point at $E= 15 \GeV$ the sub-set consists of $\{15, 20, 25, 30, 35 \GeVx \}$
%and for the point  $E= 100 \GeV$ the sub-set consists of $\{80, 90, 100, 120 \GeVx \}$.
%For each sub-set, the slope of a straight line is calculated. 
%The slope
%is $8  \cdot  10^{-5} \; \GeVinv$ at $E= 15 \GeVx$, decreases slowly towards higher energies
%and becomes $-2  \cdot 10^{-5} \; \GeVinv$ at $E= 70 \GeV$ and then increases to 
%$-1  \cdot 10^{-7}   \; \GeVinv$ at $E= 120 \GeVx$.
%
%The local non-linearity is always smaller than $ 10^{-4} \; \GeVinvx$.
%
The result for the default measurement and for the systematic variations
added in quadrature\footnote{Here, we only consider the systematics which are not related to the 
uncertainty of the test-beam geometry, 
i.e., normalisation of the presampler, the strips and the middle, the timing, the fit range, the lateral extension of the shower
and using the different Monte Carlo simulations to extract the calibration constants.}
at each energy point (where the slope can be calculated) is shown
in Fig.~\ref{fig:localslopes}. In the region relevant for the measurement
of the $\W^\pm$-mass %, i.e. around $40$\GeVx, 
the local slope is 
known to a level of about $\pm 4 \cdot 10^{-5}\GeVinvx$. This translates
roughly to an uncertainty of $15$\MeV on the $\W^\pm$-mass.

%%%%%%%%%%%%%%%%%%%%%%%%%%%%%%%%%%%%%%%%%%%%%%%%%%%%%%%%%
%%%%%%%%%%%%%%%%%%%%%%%%%%%%%%%%%%%%%%%%%%%%%%%%%%%%%%%%%%%%%%%%%g

\subsection{The Resolution Results}
\label{sec:resresults}
\begin{figure}[th]
\begin{center}
\psfig{figure=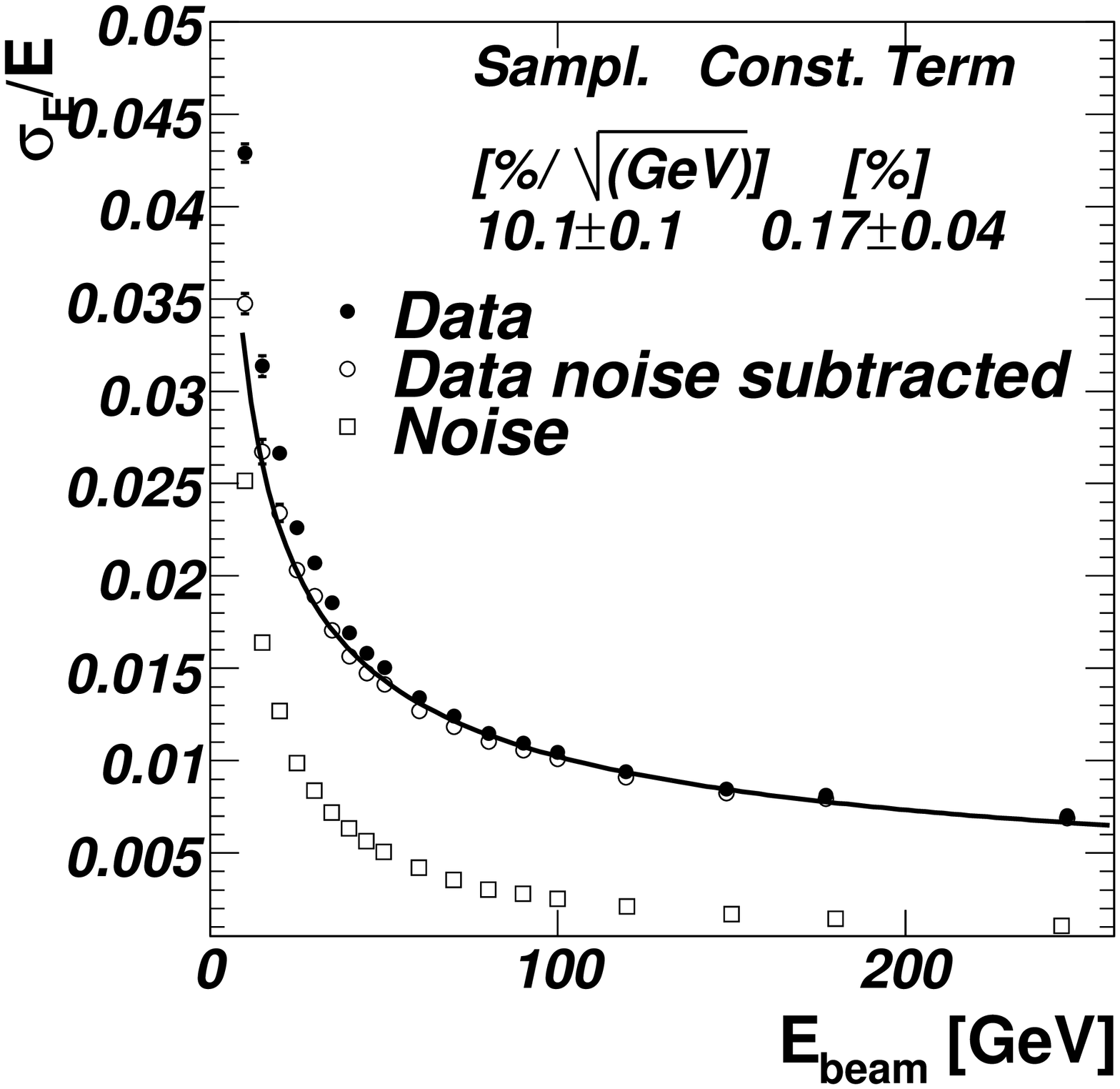,
bbllx=0,bblly=0,bburx=560,bbury=630,angle=270,clip,
width=14.5cm}
\end{center}
%\vspace{-0.5cm}
\caption{Fractional energy resolution  as a function of the beam energy.
Shown are the data before (closed circles) and after (open circle)
the gain dependent noise subtraction. 
Overlayed as a line is 
%the resolution in the Monte Carlo simulation and 
a parameterisation of the resolution based on eq.~\ref{eq:resformula} obtained from a fit.
The open squares indicate the subtracted noise contribution.
\label{fig:eres}}
\end{figure}

\noindent 
The energy resolution is obtained from the standard deviation of the Gaussian fit described 
in section~\ref{sec:linresults}.
The relative resolution %, i.e. the standard deviation divided by the mean reconstructed energy, 
as a function of the electron beam energy
is  shown as closed circles in Fig.~\ref{fig:eres}. 
%Superimposed as line are the results from the Monte Carlo simulation.
%The resolution in the data is slightly better than the resolution in the Monte Carlo simulation.

Since the noise depends on the electronic gain of the cells, the noise
is subtracted for each energy point to obtain
the intrinsic resolution of the calorimeter. 
The noise is evaluated as described in section~\ref{sec:g4}. 
%The reduction thanks to the use of the optimal
%filtering coefficients has been taken into account.
The noise is about $250$\MeV and slightly increases towards higher energies.
The noise contribution to the resolution is shown in Fig.~\ref{fig:eres} as open squares.
The data where the noise contribution and in addition the beam spread has been subtracted
are shown in Fig.~\ref{fig:eres} as open circles.
A function of the following form is fitted:
\begin{eqnarray}
\frac{\sigma_E}{E} = \frac{a}{\sqrt{E}} \oplus b,
\; {\rm with } \;
a = 10.1 \pm 0.1 \% \cdot \sqrt{\GeV}
 \; {\rm and} \;
b = 0.17 \pm 0.04 \%.
\label{eq:resformula} 
\end{eqnarray}
The symbol $\oplus$ indicates that the two terms are added in quadrature.
%The term $a$ is the stochastic term due to the sampling fluctuations and $b$ is the
%constant term reflecting local non-uniformities in the calorimeter.
%
The quoted errors are only statistical.
The fit function is overlayed to the data points % as solid line. 
and gives a good description of the
energy dependence in the data.
%
%Given the relatively large $\eta$ of the impinging electron,
The result is compatible with previous test-beam results taken in this $\eta$ region \cite{modul0}.

%%%%%%%%%%%%%%%%%%%%%%%%%%%%%%%%%%%%%%%%%%%%%%%%%%%%%%%%%%%%%%%%%
%%%%%%%%%%%%%%%%%%%%%%%%%%%%%%%%%%%%%%%%%%%%%%%%%%%%%%%%%%%%%%%%%g

\section*{Conclusions}
\label{sec:conclusions}
\noindent 
Electron energy measurements with a module of the ATLAS electromagnetic barrel \LAr calorimeter
have been studied in the range from $10$ to $245$\GeV
impinging at $\eta=0.687$
at the CERN H8 test-beam upgraded for precision momentum measurement.
%with an upgrading system measuring the electron beam energy.
The beam energy has been monitored in the range $10< E < 180$\GeV
with an accuracy of
$3  \cdot 10^{-4}$ for the uncorrelated uncertainty and an uncertainty of 
$11$\MeV common to all energy points.

A calibration scheme has been developed for electrons that provides a good linearity
and a good resolution at the same time. 
The mean reconstructed energy, the energy distributions, as well as the longitudinal
and lateral energy profiles, are well described by the Monte Carlo simulation.
Based on this simulation the data have been corrected for various effects
involving 
the intrinsic non-linearities due to the varying sampling fraction,
energy losses due to upstream and downstream interactions, energy depositions
outside the electron cluster, and losses due to Bremsstrahlung at the
beginning of the beam line.

In the energy range $15 \le E \le 180$\GeVx, the reconstructed energy response is linear 
within $\pm 0.1$\%. The point at $E = 10$\GeV
is  about $0.7\%$ lower than the other beam energies.
At $E = 245$\GeV no precise beam energy measurement was available.

The systematic uncertainties due to the limited knowledge of the
test-beam or detector set-up or due to reconstruction effects is
generally larger at low energies (up to about $\pm 0.1$\%), 
but negligible at high energies.
The non-linearity observed in the energy range of $40$\GeV and higher matches,
if extendable to the whole calorimeter, the requirements for the
$W^\pm$-mass measurement aiming for a precision of $15$\MeVx.

The sampling term of the energy resolution is found to be $10 \% \cdot \sqrt{\GeV}$, the
local constant term is $0.17$\%.

\section*{Acknowledgements}
We would like to thank D. Cornuet, J. Dutour from the CERN AT/ME department
and Y. Gaillard and J.P. Brunet from the CERN AB/PO department for valuable
help on setting up the beam.
We thank our NIKHEF ATLAS colleagues A.~Linde for kindly supplying the Hall probes used for the precise
beam energy determination and H.~Boterenbrood for help with the read-out.
We are indebted to our technicians and engineers for their
contribution to the construction
and running of the calorimeter modules and the electronics.
We would like to thank the accelerator division for the good working
conditions in the H8 beam-line.
Those of us from non-member states
wish to thank CERN for its hospitality.
\bibliography{nim}
\end{document}

%% file: abbrevia.tex
\newcommand{\W}{\mbox{$W$}}

\newcommand{\ns}{\mbox{\rm ~ns}~}
\newcommand{\nsx}{\mbox{\rm ~ns}}

\newcommand{\cm}{\mbox{\rm ~cm}~}
\newcommand{\cmx}{\mbox{\rm ~cm}}
\newcommand{\mm}{\mbox{\rm ~mm}~}
\newcommand{\mmx}{\mbox{\rm ~mm}}

\newcommand{\TeVx}{\mbox{\rm ~TeV}}
\newcommand{\GeV}{\mbox{\rm ~GeV}~}
\newcommand{\GeVx}{\mbox{\rm ~GeV}}
\newcommand{\GeVinv} {\mbox{\rm ~GeV$^{-1}$}}
\newcommand{\GeVinvx}{\mbox{\rm ~GeV$^{-1}$}}
\newcommand{\MeV}{\mbox{\rm ~MeV}~}
\newcommand{\MeVx}{\mbox{\rm ~MeV}}

\newcommand{\standclus}{\mbox{$3{\rm x}3$}}

\newcommand{\Geant}{\mbox{\rm GEANT4~}}

\newcommand{\LAr}{\mbox{\rm LAr}~}
\newcommand{\LArx}{\mbox{\rm LAr}}
\newcommand{\Xzero}{\mbox{$X_0$}~}
\newcommand{\Xzerox}{\mbox{$X_0$}}

\newcommand{\fsample}{\mbox{$f_{\rm samp}^{\rm e}$}~}
\newcommand{\fsamplex}{\mbox{$f_{\rm samp}^{\rm e}$}}

%% file: table.tex
\begin{table}
\begin{tabular}{|c|c|c|c|c|c|c|} 
\hline
events   & nom. E           & current    &  $\int Bdl$           &  magnet  & syn. rad. corr. & E   \\
         & [\GeV]           & [{\rm A}]  &  [{\rm T $\cdot$ m}]          &  corr.   &   [\GeV]        & [\GeV]   \\ 
\hline
$6703$   &  $10.$           &   $62.315$ &  $1.3664$ & $0.99975$ & $0.0000$ & $10.092 \pm 0.013$  \\  
\hline
$2727$   &  $15.$           &   $93.133$ &  $2.0379$ & $0.99975$ & $0.0000$ & $15.041 \pm 0.009$  \\ 
\hline
$6463$   &  $20.$           &  $124.363$ &  $2.7184$ & $0.99975$ & $0.0001$ & $20.061 \pm 0.010$  \\ 
\hline
$6746$   &  $25.$           &  $155.513$ &  $3.3980$ & $0.99975$ & $0.0003$ & $25.074 \pm 0.011$  \\
\hline
$8403$   &  $30.$           &  $186.689$ &  $4.0789$ & $0.99975$ & $0.0006$ & $30.098 \pm 0.012$  \\
\hline
$7842$   &  $35.$           &  $217.900$ &  $4.7607$ & $0.99975$ & $0.0012$ & $35.128 \pm 0.013$  \\
\hline
$8791$   &  $40.$           &  $248.973$ &  $5.4397$ & $0.99975$ & $0.0020$ & $40.137 \pm 0.014$  \\
\hline
$8314$  &  $45.$            &  $280.075$ &  $6.1201$ & $0.99975$ & $0.0032$ & $45.148 \pm 0.014$        \\
\hline
$18366$  &  $50.$           &  $311.270$ &  $6.8005$ & $0.99975$ & $0.0049$ & $50.174 \pm 0.017$  \\
\hline
$9465$   &  $60.$           &  $373.484$ &  $8.1596$ & $0.99975$ & $0.0102$ & $60.197 \pm 0.020$  \\
\hline
$9507$   &  $70.$           &  $435.705$ &  $9.5187$ & $0.99975$ & $0.0189$ & $70.216 \pm 0.022$  \\
\hline
$9829$   &  $80.$           &  $497.673$ & $10.8715$ & $0.99976$ & $0.0322$ & $80.184 \pm 0.025$  \\
\hline
$9973$   &  $90.$           &  $559.524$ & $12.2208$ & $0.99976$ & $0.0515$ & $90.120 \pm 0.028$  \\
\hline
$22685$  & $100.$           &  $621.128$ & $13.5634$ & $0.99976$ & $0.0781$ & $100.000 \pm 0.033$ \\
\hline
$9585$   & $120.$           &  $744.113$ & $16.2383$ & $0.99980$ & $0.1605$ & $119.658 \pm 0.037$ \\
\hline
$9367$   & $150.$           &  $926.708$ & $20.1866$ & $0.99995$ & $0.3806$ & $148.594 \pm 0.046$ \\ %0.3836-0.003
\hline
$4960$   & $180.$           & $1115.230$ & $24.1095$ & $1.00011$ & $0.7698$ & $177.182 \pm 0.056$ \\ %0.7818-0.012
%\hline
%$42780$   & $245.$           & ---        & ---       & ---       & ---     & --- \\
\hline
\end{tabular}
\vspace{0.5cm}
\caption{Summary of the used data sets.
Number of the selected electron events, nominal beam energy, induced current in the
magnetic field, magnet  bending power, corrections for magnets and synchrotron
radiation and the measured beam energy. 
The overall beam momentum scale is normalised to $100$\GeVx.
At $E=245\GeV$no precision energy measurement is possible with the used magnet set-up.
\label{tab:beam_momentum}}
\end{table}
%%%%%%%%%%%%%%%%%%%%%%%%%%%%%%%%%%%%%%%%%%%%%%%%%%%%%%%%%%%%%%%%%%%%%%%%%%%%%%